\documentclass[chicago, iop, apj]{emulateapj}
\usepackage{lineno}

\usepackage{newtxtext,newtxmath}
\usepackage[T1]{fontenc}
\usepackage{array, booktabs, makecell, bm}
\usepackage[colorlinks=true, allcolors=blue]{hyperref}

\usepackage{amsmath}
\usepackage{graphicx}
\usepackage{xcolor}

\shortauthors{Authors}

\begin{document}

\title{The Radio and Microwave Sky as Seen by Juno on its Mission to Jupiter}
\author{
C. J. Anderson\altaffilmark{1}, P. Berger\altaffilmark{2}, T.-C. Chang\altaffilmark{1,3}, O. Dor\'e\altaffilmark{1,3}, S. Brown\altaffilmark{1}, S. Levin\altaffilmark{1}, M. Seiffert\altaffilmark{1}}

\email{christopher.anderson@jpl.nasa.gov}

\altaffiltext{1}{Jet Propulsion Laboratory, 4800 Oak Grove Dr., Pasadena, CA 91011, USA}
\altaffiltext{2}{Hinge Health, 455 Market Street, Suite 700, San Francisco, CA 94105, USA}
\altaffiltext{3}{California Institute of Technology, 1200 California Blvd., Pasadena, CA 91125, USA}

\shortauthors{Anderson et al.}
\shorttitle{Juno All-Sky Maps}

\begin{abstract}
We present six nearly full-sky maps made from data taken by radiometers on the Juno satellite during its 5-year flight to Jupiter. The maps represent integrated emission over $\sim 4\%$ passbands spaced approximately in octaves between 600 MHz and 21.9 GHz. Long time-scale offset drifts are removed in all bands, and, for the two lowest frequency bands, gain drifts are also removed from the maps via a self-calibration algorithm similar to the NPIPE pipeline used by the Planck collaboration. We show that, after this solution is applied, statistical noise in the maps is consistent with thermal radiometer noise and expected levels of correlated noise on the gain and noise drift solutions. We verify our map solutions with several consistency tests and end-to-end simulations. We also estimate the level of systematic pixelization noise and polarization leakage via simulations.
\end{abstract}

\section{Introduction}

Sky maps of diffuse emission across the electromagnetic spectrum have been invaluable to our understanding of both extragalactic and Milky Way astrophysics.
All-sky microwave maps of the relic thermal cosmic microwave background radiation from the early universe have been the most important data sets in fleshing out the current $\Lambda$CDM model for the history of the universe \citep{fixsen1996cosmic, smoot1992structure, spergel2003first, ade2016planck}. At infrared wavelengths, all-sky maps have measured both Galactic star formation \citep{dwek1995morphology} and extragalactic star formation history \citep{dwek1998tentative, finkbeiner2000detection, lenz2019large, serra2014cross}. At radio frequencies, maps are dominated by Galactic synchrotron emission, with a significant contribution from extragalactic synchrotron point sources. These radio maps serve as important tools for modeling and removing both total intensity and polarized foregrounds in CMB maps \citep{ichiki2014cmb, akrami2020planck, rubino2023quijote} and in 21-cm intensity maps \citep{bernardi2009foregrounds, shaw2014all}. They have also been used to study magnetic field and Galactic halo properties of the Milky Way \citep{orlando2013galactic}. Bright extragalactic point sources can be isolated from Galactic emission (e.g. the NVSS catalog, \cite{nvss}), but it is difficult to distinguish between dim extragalactic radio sources and Milky Way synchrotron emission. A separation of Galactic and extragalactic emission was attempted by the ARCADE2 team, which measured the absolute temperature of a large chunk of the sky over a range of frequencies from 3 GHz to 90 GHz \citep{2011ApJ...734....5F}. Comparing their absolute temperature with that reported by lower frequency surveys \citep{haslam1982408, maeda199945, roger1999radio}, they found that all the data is consistent with, in addition to the CMB monopole, a bright radio background with a synchrotron-like spectral index. They interpret this background as extragalactic in origin \citep{kogut2011arcade} due to its excess above both a co-secant Galactic synchrotron model and a Galactic synchrotron template based on the square of the Milky Way [CII] emission map from the FIRAS instrument \citep{1999ApJ...526..207F}. The origin of such an extragalactic radio excess remains unclear, as it is much brighter than extrapolations of known point source populations \citep{2010MNRAS.409.1172S}.

In this work, we present six new nearly all-sky maps made from radiometers on the Juno spacecraft during its 5-year cruise from Earth to Jupiter. The frequency bands are spaced in octaves over a range from 0.6 GHz to 21.9 GHz. 
To the authors' knowledge, these are the lowest frequency nearly all-sky maps created by a single instrument. Although the angular resolution of the maps is low, they provide well-characterized measurements of the large-scale angular structure of the sky over several important frequency bands that are lacking in all-sky maps \citep{de2008model, zheng2017improved}. The lower frequency maps will help characterize spatial and spectral variations of synchrotron emission. The higher frequency maps clearly show the presence of the Doppler-shifted CMB dipole, and they contain significant emission from the anomalous microwave background radiation (AME) (see \cite{dickinson2018state} for a review). This work constitutes the first demonstration of on-sky self-calibration for low-frequency, spinning spacecraft, full-sky surveys. We validate our method and noise model with end-to-end simulations and residual tests. 

The organization of this paper is as follows. Section \ref{section:Instrument} summarizes essential details of Juno's flight, scan strategy, microwave receiver systems, data calibration technique, and polarized response. Section \ref{section:map_and_errors} describes the map-making formalism used to remove long time-scale additive noise drifts and multiplicative gain drifts. It also derives forms for the statistical errors of these maps and summarizes estimated sources of systematic errors. Section \ref{section:results} presents the resulting maps, as well as plots of statistical and systematic error estimates. It also presents verification tests performed via residual noise analyses and time-stream simulations. Section \ref{section:discussion} concludes with a discussion of the Juno maps in the context of existing full-sky measurements and physical emission models. 
The maps from this analysis, along with estimates of systematic and statistical errors, are publicly available. The link can be found in section \ref{section:data_availability}.

\section{Juno Observations of the Microwave Sky}\label{section:Instrument}
\cite{janssenetal} contains detailed information on the Juno Microwave Radiometer (MWR) design, instrument, and calibration. We provide here a summary of the details relevant to the cruise analysis.

\subsection{Juno's Cruise Mission to Jupiter}
The Juno Cruise Mission spanned a $\sim$5 year period from launch in late 2011 to Jupiter Orbit Insertion (JOI) in 2017. This period consisted of a two-year inner cruise phase, an Earth flyby, and a subsequent three-year coast to Jupiter (the outer cruise). During the cruise, Juno completed about one and a half orbits around the Sun. The spacecraft spun continuously about an axis that aligned with the high-gain communication antenna, which is perpendicular to the pointing axes of the MWR antennas. The antenna beams, therefore, swept through great circles on the sky as the spacecraft rotated. Since the communication antenna was kept in alignment with the Earth, Juno's solar orbit during the cruise allowed the MWR to observe nearly the full sky. The spin rate varied between 1 and 2 rpm throughout the cruise: it started at 1 rpm, then switched to 2 rpm mid-way through 2012, switched back to 1 rpm for the end of 2013 through the start of 2015, and then went back to 2 rpm for the remainder of the cruise.

The Juno Cruise Mission data is publicly available on the Planetary Data System (PDS) \citep{Juno_Cruise_PDS}. Data for the 5-year cruise mission is stored in  ASCII CSV format in files labeled by the hour and in directories for the day and year. Geometric information on the spacecraft's orientation is stored in Geometric Data Record (GRDR) files including pointing and polarization angle information. Instrument information is stored in Instrument Data Record (IRDR) files, including radiometric data and on-board thermometry for each sub-system. The maps created in this work use the calibrated antenna temperatures from the PDS data taken between 2012 and mid-2016. The calibrated PDS data is found under columns [`R1\_1TA', `R2\_1TA', `R3TA', `R4TA', `R5TA', `R6TA'] for each of the six receivers. The calibration procedure that was used to generate this PDS data is described in section \ref{subsection:on_board_cal}. We make some cuts to this data to remove RFI and periods of relatively unstable receiver temperature and gain, as described in section \ref{subsection:data_cuts}.

\subsection{Microwave Receiver Systems}

The Juno spacecraft has a hexagonal body, to three sides of which are mounted extended solar panels. Two of the three remaining sides are covered by the 6 MWR antennas. The receivers are sensitive to $\sim 4\%$ passbands spaced approximately in octaves between 600 MHz and 21.9 GHz. The central frequency of each receiver is 0.600, 1.248, 2.597, 5.215, 10.004, and 21.900 GHz, for receivers R1 to R6, respectively. The angular resolution of each antenna, numbered correspondingly A1 to A6, is limited by the physical size of the spacecraft. These sizes were chosen to give roughly consistent angular resolution across the band. One side is therefore taken up entirely by A1, and the remaining 5 antennas cover another one of the hexagon's faces, pointing 120$^\circ$ away. A1 through A5 are slotted waveguide antenna designs and A6 is a corrugated horn. All are singly linearly polarised with E-field directions parallel to the spacecraft spin axis. More details on the MWR system are included in table \ref{table1}.

The MWR integrates the input noise power from each receiver in 100 ms samples. A sample can observe either the pure antenna input, the antenna combined with noise diode input, or the internal reference load alone. Typically, the noise diode and reference will be observed once in a 1 s period, leading to an $80\%$ on-sky duty cycle. However, during the cruise, the MWR was operated at a reduced data rate with 10 s pauses between each 1 s observation burst.

\begin{table*}\label{table1}
\centering
\tabcolsep=0.11cm
\setlength\tabcolsep{0pt}
\begin{tabular*}{\textwidth}{@{\extracolsep{\fill}} c c c c c c }
\toprule
 Receiver & \thead{Nominal centre \\ frequency [GHz]} & Bandwidth [MHz] & Beam FWHM [$^\circ$] & \thead{Nominal NEDT \\  $\left[Ks^{1/2}\right]$}  & $\sigma_N$ $[K]$ \\ 
 \midrule
 R1 & 0.600 & 21 & 19.7 & 0.187 & 0.59  \\  
 R2 & 1.248 & 43.75 & 19.8 & 0.171 & 0.54   \\
 R3 & 2.597 & 84.5 & 11.9 & 0.133 & 0.42  \\
 R4 & 5.215 & 169 & 11.9 & 0.123 & 0.39  \\
 R5 & 10.004 & 325 & 11.9 & 0.066 & 0.21  \\
 R6 & 21.900 & 770 & 10.7 & 0.060 & 0.19   \\
 \bottomrule 
\end{tabular*}
\caption{Summary of the receiver systems for the 6 Juno bands, labeled R1 through R6. The fifth column shows the nominal NEDT \citep{janssenetal}, which is the expected noise in a 1-second integration time. The last column converts this to the expected noise level in the time-stream, $\sigma_N$, where the integration time is 0.1 seconds.}
\end{table*}

\subsection{On-Board Calibration}\label{subsection:on_board_cal}
Each of Juno's 6 MWR systems is calibrated via Dicke-switching measurements of an internal load and by injections of power from a calibration noise diode. During the cruise, Juno measures power from the sky in time bins of 0.1 seconds. Once during every 1-second regular observation cycle, the MWR injects power from the noise diode into a regular 0.1-second sky observation. Also once per cycle, the MWR switches to measure the internal reference load for 0.1 seconds. A running average of measurements of this reference load is subtracted from the sky data to remove the receiver temperature, $T_{RX}$, and any DC offset from the data stream. A running average of the noise diode power is used to calibrate the system gain. Mathematically, the final calibrated antenna temperature is given by
\begin{linenomath*}\begin{equation}\label{eq:calibration}
T_A(t) = \frac{C_A(t) - \overline{C_R}(\tau_R)}{\overline{G_{\rm sys}}(\tau_G)} + \overline{T_{\rm off}}(\tau_R), 
\end{equation}\end{linenomath*}
where $C_A(t)$ is the output from the Analog-to-Digital Converter (ADC) when the system is observing the sky; $\overline{C_R}(\tau_R)$ is a running average of the ADC output when the system is observing the internal load, computed over a several minute time-scale, $\tau_R$; $\overline{G_{\rm sys}}(\tau_G)$ is a running average of the system gain estimate, computed over a several minute time-scale, $\tau_G$; and $\overline{T_{\rm off}}(\tau_R)$ is a running average estimate of the power in the internal calibration load multiplied by internal reflection coefficients, computed over a several minute time-scale, $(\tau_R)$. The system gain estimate comes from a running average of differences in antenna temperature with the noise diode on and off:
\begin{linenomath*}\begin{equation}\label{eq:gain_estimate}
\overline{G_{\rm sys}}(\tau_G) = \left< \frac{C_{A+ND} - C_A}{T_{ND}}\right>_{\tau_G}
\end{equation}\end{linenomath*}
Any drifts in the noise temperature of the internal load that are not accounted for in $\overline{T_{\rm off}}(\tau_R)$ will result in an overall offset on $T_A(t)$. Unmeasured drifts in the noise diode temperature can lead to an incorrect gain for $T_A(t)$. This work therefore uses a map-making formalism, described in sections \ref{subsection:noise_drifts} and \ref{subsection:gain_drifts}, that attempts to correct for long time-scale drifts in the offset and the gain.

\subsection{Data Cuts}\label{subsection:data_cuts}
We cut approximately 12.5\% of the time-stream data from our final map-making analysis via a procedure designed to remove RFI and to find periods of integration where the calibrated receiver temperature offset and gain vary slowly compared to our planned 4-hour de-striping periods. First, to remove strongly RFI-contaminated data, we fit the $\sim$80 million data points for each receiver to a third order polynomial. We then flag and remove any data points that are more than 12 median absolute deviations (MADs) from the polynomial fit in any of the 6 bands. This results in the removal of a very small percentage of highly RFI-contaminated data points, detected mainly via the 600 MHz receiver. Upon examining the remaining data, we find that it sometimes changes dramatically at the edges of gaps between the continuous several month long chunks of time when the MWR was collecting data. We cut a portion of the data on these edges, based on time-stream residuals that are computed by subtracting a preliminary map from the data. These residuals are fit to a noise model consisting of diagonal thermal noise and $1/f$ noise. Periods on the edge of integration chunks with high $1/f$ noise are removed until all the remaining data periods have a $1/f$ knee frequency longer than a day. 

\subsection{Polarized Response}

\begin{figure*}
    \includegraphics[width=\textwidth]{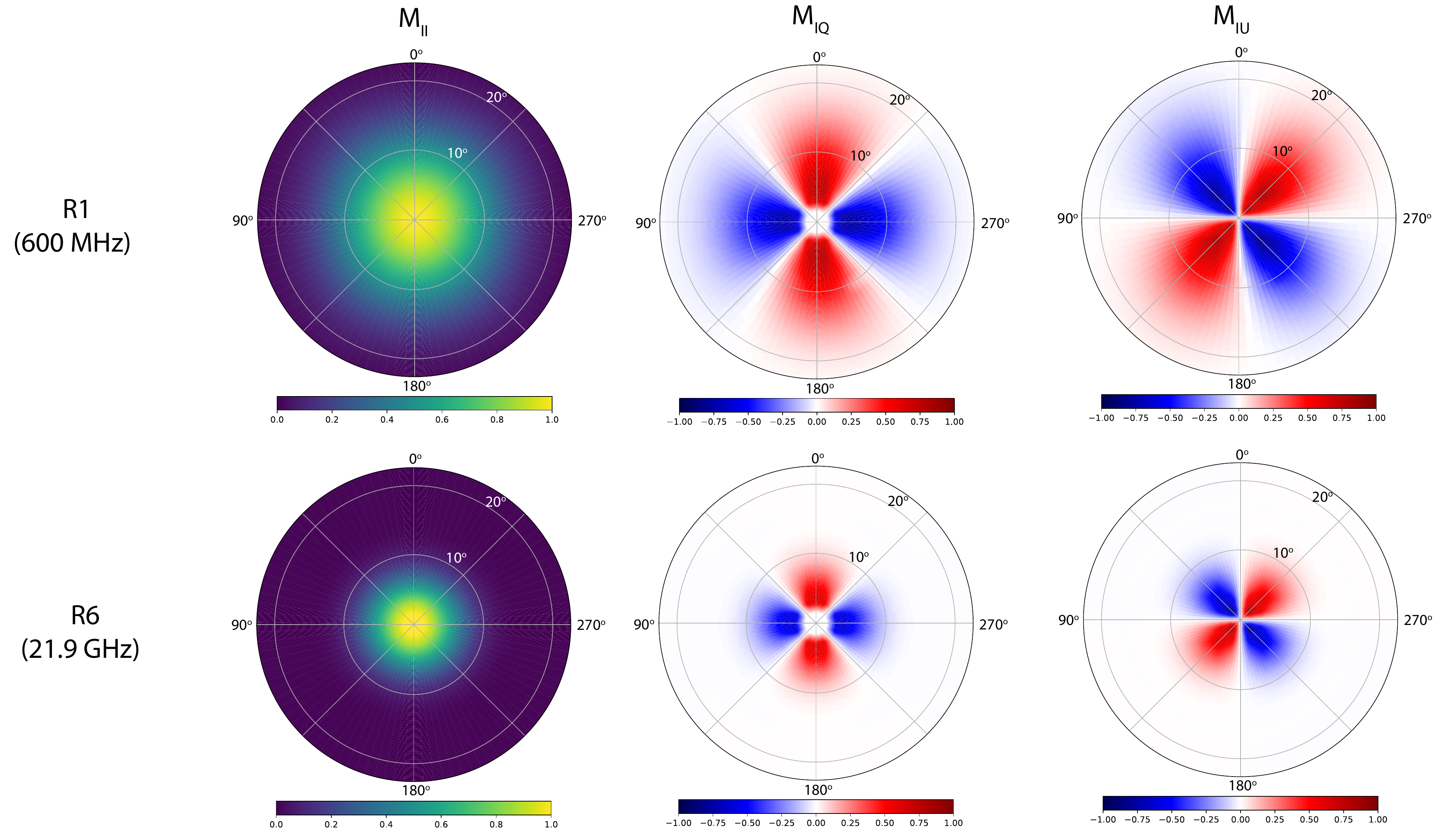}
    \caption{ \label{fig:polarized_beams} The computed beam responses for receivers R1 and R6. The plots show the Mueller matrix responses to pure Stokes $I$ (left, $m_{II}$), pure Stokes $Q$ (middle, $m_{IQ}$), and pure stokes $U$ (right, $m_{IU}$). All the responses are normalized relative to the stokes $I$ response at the beam center. The polar coordinate indicates the polar angle ($\theta$) from the beam center. The azimuthal angle ($\phi$) is defined with respect to the receiver polarization axis, with up ($\phi=0^{\circ}$) being parallel to the polarization axis. $Q$ polarization is defined to be positive along the $\pm\hat{\theta}$ direction and negative in the azimuthal ($\pm\hat{\phi}$) direction, while the plus and minus $U$ directions are 45 degrees rotated from these. The polarized responses broadly follow the expected behavior for a vertically polarized antenna, showing a strong positive response when the polarization source is aligned with the antenna's polarization axis and a strong negative response when the source's polarization is orthogonal to the receiver's polarization axis. One odd feature is the hole at the center of the $m_{IQ}$ patterns. This feature is not physical but is a by-product of the Healpix pixelization scheme, where only the azimuthal angles of $45^{\circ}$, $135^{\circ}$, $225^{\circ}$, and $315^{\circ}$ are defined on the ring with the smallest polar angle. The plots shown here are created from interpolations of Healpix beam maps at $N_{side}=32$. Increasing the Healpix resolution can shrink the size of the central hole in $m_{IQ}$. Judging that they are accurate enough for estimating the polarized contribution to our maps, we use $N_{side}=32$ for polarization leakage simulations (section \ref{subsection:leakage_sim}).}
\end{figure*}

Each of the six Juno receivers measures only one linear polarization.
This is modeled mathematically as the receiver measuring only the Stokes $I$ parameter, but with significant leakage from $Q$ or $U$, depending on the rotation angle of the receiver. Schematically,
\begin{linenomath*}\begin{equation}
\hat{I} = m_{II}I + m_{IQ}Q + m_{IU}U,
\end{equation}\end{linenomath*}
where $I$, $Q$, and $U$ are the true polarized sky at a particular pointing, $\hat{I}$ is the estimate of the sky obtained from treating the power from the Juno receiver as a measurement of stokes $I$, and where $m_{II}$, $m_{IQ}$, and $m_{IU}$ are Mueller matrix elements describing the contributions to $\hat{I}$ from $I$, $Q$, and $U$. In reality, these Mueller matrix elements are not single numbers, but patterns defined over all polar angles relative to the beam center and all azimuthal angles relative to the receiver polarization axis.
On-ground measurements of the polarized receiver radiation patterns were performed before Juno launched \citep{janssenetal} and are available on the Juno microwave PDS \footnote{\url{https://pds-atmospheres.nmsu.edu/data_and_services/atmospheres_data/JUNO/microwave.html}}. From these measurements, we compute receiver leakage patterns. The beam response to purely un-polarized emission, $m_{II}$, is well-fit in all 6 channels by a Gaussian beam model with Full-Width-at-Half-Maximum (FWHM) shown in table \ref{table1}. Figure \ref{fig:polarized_beams} shows the Mueller matrix patterns for receivers R6 and R1. Their features are consistent with the strong polarization leakage expected from a singly linearly polarized antenna. The other frequencies are all extremely similar, with their size scaling with the beam widths. 

For the actual Juno observations, there are three relevant angles. $\theta$ and $\phi$ describe the pointing position of the receiver's beam-center in Galactic coordinates. The angle $\psi$ describes the counterclockwise rotation of the receiver's vertical polarization axis from the Galactic-coordinate-defined $-\hat{\theta}$ direction, which points towards the Galactic North Pole. At each pointing, the receiver sees the sky integrated over the pointed and rotated beam patterns.
\begin{linenomath*}\begin{equation}
\begin{split}
\hat{I}(\theta, \phi, \psi) = \int R(\theta, \phi, \psi)[m_{II}(\theta', \phi')]I(\theta', \phi')d\Omega'\\+ \int R(\theta, \phi, \psi)[m_{IQ}(\theta', \phi', \psi)]Q(\theta', \phi')d\Omega' \\+ \int R(\theta, \phi, \psi)[m_{IU}(\theta' , \phi']U(\theta', \phi')d\Omega'.
\end{split}
\end{equation}\end{linenomath*}
In the above equation,  $R(\theta, \phi, \psi)[]$ denotes the operation of pointing the beam in the $\theta, \phi$ Galactic direction and rotating the beam by an angle $\psi$ from Galactic North.
The $\psi$-dependence of the rotated $m_{IQ}$ and $m_{IU}$ beam patterns mean that pointings at the same Galactic position but different rotation angles contain different amounts of leakage from $I$ and $Q$. With sufficient rotation angle variation, it would be feasible to use this variation to solve for $I$, $Q$, and $U$ maps, though this would require computationally expensive polarized beam deconvolution. However, because the Juno polarization axis is always approximately parallel to the ecliptic plane, the beam rotation angle only varies significantly at the ecliptic poles and little elsewhere in the scan. This means that for most of the map, to vary $\psi$ enough to break the degeneracy between $I$, $Q$, and $U$, one would have to rely on small polarized contributions far from the beam center. We instead opt to treat the data as coming solely from unpolarized emission and solve only for beam-convolved $I$ maps. In section \ref{subsection:leakage_sim}, we simulate an estimate of the amount of polarized contamination present in our maps and show that it is small, thanks to beam depolarization in all bands and bandwidth depolarization \citep{2023MNRAS.520.4822F} of Faraday rotated emission in the lower frequency bands.

\section{Map-Making and Error Estimation}\label{section:map_and_errors}
This section describes the technique used to convert time-stream data to maps and outlines the process for estimating the associated errors.
Section \ref{subsection:noise_drifts} describes the de-striping technique for removing the most significant source of correlated noise: long-time-scale additive noise drifts. Section \ref{subsection:gain_drifts} describes an extension of this technique to remove long time-scale multiplicative gain drifts.  Gain drifts are successfully removed for bands R1 and R2, but the higher frequencies lack the necessary signal-to-noise ratio. Section \ref{subsection:map_resolution} describes the choices for map pixel size and scan period length. Section \ref{subsection:covariance_equations} describes the technique for estimating statistical noise. Section \ref{subsection:systematic_error_list} lists estimated sources of systematic errors.
\subsection{De-striping with the MADAM Formalism}\label{subsection:noise_drifts}
The map-making formalism used in this work is based on an extension of the MADAM de-striping technique \citep{2004A&A...428..287K, 2005MNRAS.360..390K}. The MADAM formalism assumes that the time-stream data, $\bm{y}$, consists of the pointing matrix applied to the map, $\bm{P}\bm{m}$, plus long time-scale correlated noise drifts, $\bm{n}_{corr}$, and standard diagonal radiometric thermal noise, $\bm{n}$. 
\begin{linenomath*}\begin{equation}
\bm{y} = \bm{P}\bm{m} + \bm{n}_{corr} + \bm{n}.
\end{equation}\end{linenomath*}
The time-stream data is then divided into periods, chosen to be short enough that the long time-scale noise offset is roughly constant within a period. Then, both the map, $\bm{m}$, and the amplitude of the long time-scale noise offsets, $\bm{a}$, can be treated as vectors to be solved for.
\begin{linenomath*}\begin{equation}\label{eq:MADAM_time-stream}
\bm{y} = \bm{P}\bm{m} + \bm{F}\bm{a} + \bm{n},
\end{equation}\end{linenomath*}
 where $F_{i,j}=1$ if time $i$ is in period $j$, and zero otherwise. For a sufficiently overlapping scan, a unique minimum chi-squared solution can be found, up to a degeneracy between the map monopole and a common additive offset for all the long time-scale noise drift amplitudes. The minimum chi-squared solution for $\bm{a}$ is given by
\begin{linenomath*}\begin{equation}\label{eq:general_a_solution}
 \left[ \bm{F}^T \bm{C}_n^{-1}\bm{Z} \bm{F} + \bm{C}_a^{-1} \right] \hat{\bm{a}} = \bm{F}^T \bm{C}_n^{-1}\bm{Z}\bm{y},
\end{equation}\end{linenomath*}
where $\bm{C}_n$ is the radiometric thermal noise covariance, $\bm{C}_a$ is any assumed prior on the covariance of the correlated noise drift amplitudes, and where
\begin{linenomath*}\begin{equation}
    \bm{Z} = \bm{I} - \bm{P}(\bm{P}^T\bm{C}_{n}^{-1}\bm{P})^{-1}\bm{P}^T\bm{C}_{n}^{-1}.
\end{equation}\end{linenomath*}
The solution for the map is then
\begin{linenomath*}\begin{equation}
    \hat{\bm{m}} = (\bm{P}^T\bm{C}_{n}^{-1}\bm{P})^{-1}\bm{P}^T\bm{C}_{n}^{-1}(\bm{y} - \bm{F}\hat{\bm{a}}).
\end{equation}\end{linenomath*}
In this analysis, the thermal noise is assumed to be constant and diagonal ($\bm{C}_{n} = \sigma_N^2\bm{I}$), which simplifies the equations considerably. In that case, the map solution is given by
\begin{linenomath*}\begin{equation}\label{eq:map_solution}
    \hat{\bm{m}} = (\bm{P}^T\bm{P})^{-1}\bm{P}^T(\bm{y} - \bm{F}\hat{\bm{a}}),
\end{equation}\end{linenomath*}
and the matrix $\bm{Z}$ is
\begin{linenomath*}\begin{equation}
    \bm{Z} = \bm{I} - \bm{P}(\bm{P}^T\bm{P})^{-1}\bm{P}^T.
\end{equation}\end{linenomath*}
The solution for $\bm{a}$ is given by
\begin{linenomath*}\begin{equation}\label{eq:a_solution}
\hat{\bm{a}} = \left[ \bm{F}^T \bm{Z} \bm{F} + \bm{1}\bm{1}^T \right]^{-1}\bm{F}^T \bm{Z}\bm{y},
\end{equation}\end{linenomath*}
where $\bm{1}$ is defined to be a vector of the same length as $\bm{a}$ with all values set to $1$. This equation assumes no prior knowledge of the covariance of the noise offset amplitudes, $\bm{C}_a$, except to enforce that their sum is zero via the $\bm{1}\bm{1}^T$ term. This resolves the degeneracy between a common noise offset and the map monopole and ensures that the above equation is invertible.

In this analysis, the map, $\bm{m}$, is the beam-convolved map. The pointing matrix is 1 at the pixel containing the direction in which the receiver is pointed and 0 at every other pixel.
\subsection{Calibrating Gain Fluctuations}\label{subsection:gain_drifts}

Following the Planck collaboration \citep{NPIPE}, this analysis extends the MADAM formalism to iteratively solve for the map, noise drifts, and gain drifts of the system. From equation \ref{eq:gain_estimate}, $\overline{G_{sys}}(\tau_G)$ represents the running average estimate of the system gain during period $\tau_G$. Let $G_{sys}(\tau_G)$ denote the true system gain during that period. Then drifts in the gain due to unmeasured changes in the diode power can be written as $\frac{G_{sys}(\tau_G)}{\overline{G_{sys}}(\tau_G)} = 1 + \delta g(\tau_G)$. Including these drifts, a model for the time-stream data is
\begin{linenomath*}\begin{equation}
\bm{y} = \bm{P}\bm{m} + \bm{\delta g} \circ \bm{P}\bm{m} + \bm{n}_{corr} + \bm{n}.
\end{equation}\end{linenomath*}

This is a formally non-linear problem since the second term contains products of two unknown terms, the map and the gain offsets. We make it a linear iterative problem by using the previous solution's map as a template.
\begin{linenomath*}\begin{equation}
\bm{y} = \bm{P}\bm{m}^{(n)} + \bm{\delta g}^{(n)} \circ \bm{P}\hat{\bm{m}}^{(n-1)} + \bm{n}_{corr}^{(n)} + \bm{n}.
\end{equation}\end{linenomath*}

Given a previous map solution, $\hat{\bm{m}}^{(n-1)}$, as a template for the sky, an optimal map, noise offset, and gain offset solution can be solved for via equations \ref{eq:a_solution} and \ref{eq:map_solution}. To include gain fluctuations, $\mathbf{F}$ must be extended with columns that are zero at times not contained within a given scan, and, for times that are within the given scan, are equal to the pointing matrix applied to the previous iteration's estimate for the (beam convolved) map: $\bm{P}\hat{\bm{m}}^{(n-1)}$. The length of the vector $\bm{a}$ doubles, with the first half representing the noise offset amplitudes and the second half representing the gain offset amplitudes. The $\bm{1}\bm{1}^T$ term in equation \ref{eq:a_solution} should be replaced with terms that add 1s to the upper left and lower right blocks of $\bm{F}^T \bm{Z} \bm{F}$. This assures that both the mean noise offset and the mean gain offset are zero. 

The map-making process begins by setting the gain offsets to zero and solving only for a sky map and noise offsets. This map then serves as the first sky template, $\hat{\bm{m}}^{(0)}$, and the gain offsets are allowed to vary in all of the following iterations. The procedure is repeated until the map, gain offsets, and noise offsets converge. Simulations show that, for the two lowest frequencies, accurate solutions to the gain and noise offsets are quickly found.  For the higher frequency bands, the Milky Way fluctuations in the map are much smaller compared to the noise, which leads to difficulties.  Although the solutions converge, simulations show that, for R3, the noisy gain solution is slightly farther from the true gain than simply assuming no gain offsets. 
For R4 through R6, gain offset solutions fluctuate rapidly and attain non-physical values less than -1 (indicating negative gain). A $\chi^2$ analysis reveals that there is a negligible improvement when using gain solutions at these frequencies compared to the noise-offset-only solution. We therefore conclude that the signal-to-noise ratio is insufficient to justify solving for gain drifts at these frequencies, and we instead report the noise offset only solutions in R3 through R6. 

\subsection{Effects of Map Resolution and Period Length}\label{subsection:map_resolution}
In choosing a resolution at which to make the maps, there is a trade-off between pixelization noise and thermal noise on the estimated noise and gain offsets. Lower resolution maps treat pointings that are looking at slightly different parts of the sky as overlapping, which can lead to pixelization errors in the final map and the derived noise and gain offsets. However, too fine of a map resolution can result in very little overlap between different scans, resulting in low constraining power and high thermal noise on the estimated noise and gain offset amplitudes. We release maps made at a resolution of $N_{side}=64$, which, at approximately 1 degree of angular resolution, provides an excellent sampling of the large Juno beams but is coarse enough to avoid numerous gaps in coverage. However, as we show in section \ref{subsection:pixelization_noise}, uneven sampling within each pixel couples with the steep gradient of Galactic emission to produce pixelization errors that are larger than the thermal noise near the bright Galactic plane for bands R1 and R2. For these bands simulations show that the best resolution to minimize errors on the gain and noise offsets is $N_{side}=512$, where the pixelization errors are below the thermal noise level. So, for R1 and R2, we compute solutions at $N_{side}=512$ and also release a re-binning of those maps to $N_{side}=64$. Although the gain and noise offset solutions for those re-binned $N_{side}=64$ maps have negligible pixelization errors, the maps themselves have large pixelization errors near the Galactic plane. Therefore, we also release simulated pixelization noise corrections that can be added to the maps to approximately remove the effect of pixelization noise. These corrections are based on applying the Juno scan strategy to a high-resolution model of the Galaxy based on the PySM sky model \citep{pysm}. 

The map-maker solves for long time-scale noise and gain drifts that are assumed to be constant within each period. One is free to choose the length of those periods. They should be short enough that noise and gain drifts are negligible over that period but not so short that solving the linear system becomes computationally impractical. We choose 4-hour periods, of which there are roughly 5000 in all the Juno cruise data, and we count on the on-board calibration procedure for stability on shorter time-scales. An analysis of the map-subtracted noise residuals (section \ref{subsec:residual_noise_analysis}) shows no unexpected correlated noise to indicate that the 4-hour period is too long. The noise residuals are dominated by the thermal diagonal component. As expected, there is also a lower level of correlated calibration noise at the $\sim$100 second time-scale due to Juno's on-board calibration procedure, which computes the gain and system temperature every few minutes, as described in section \ref{subsection:on_board_cal}. Noise residuals at time-scales between 100 seconds and 4 hours are generally positively correlated, with a magnitude 100 times smaller than the calibration noise level. This is consistent with the expected noise level on the 4-hour MADAM noise/gain amplitudes, assuming that the calibration noise dominates those measurements (equation \ref{eq:a_cov}.) 

\subsection{Statistical Errors: Map-space covariance}\label{subsection:covariance_equations}
The statistical errors of the maps are modeled with 3 terms:
\begin{enumerate}
\item \textbf{Diagonal radiometric thermal noise}. The expected thermal noise is $\frac{\sigma_N^2}{N_i} = \sigma_N^2 (\bm{P}^T\bm{P})^{-1}$, where $N_i$ is the number of time samples in pixel $i$ and $\sigma_N$ is the standard deviation of the thermal noise in the time-stream (shown in the last column of table \ref{table1}). For the value of $\sigma_N$, we use the standard deviation of the time-stream residuals, which includes contributions from both diagonal thermal noise and short time-scale correlated calibration noise. As figure \ref{fig:residual_variance} shows, the time-stream residuals are close to the expected thermal noise values. Section \ref{subsec:residual_noise_analysis} shows that this thermal noise term is a good model of statistical errors between maps made with different halves of the data and a common MADAM solution, though maps R1 and R2 have substantial pixelization noise (a systematic effect from the scan pattern, described in section \ref{subsection:systematic_error_list} and discussed in detail in section \ref{subsection:pixelization_noise}). 
\item \textbf{Correlated noise.} This term is due to the long time-scale of the noise and gain drifts and their uncertain estimation with our map-making formalism. An analytic form for these errors is provided in the text of this section. Due to minute-scale correlated calibration noise, the scale of this analytic estimate is too low, so we adjust its amplitude to match simulations (see section \ref{subsec:time-stream_sims}).
\item \textbf{Pointing errors.} Given a beam-convolved model for the true sky, $\bm{m}$, and estimated random pointing errors of magnitude $\Delta \beta$, an expression for the expected variance in the final map due to random pointing errors is 
$\frac{\left<\Delta \beta ^2 \right>}{2N_i}\left[ \left(\partial{\bm{m}}/\partial{\theta}\right)^2 + \left(\frac{\partial{\bm{m}}/\partial{\phi}}{\sin{\theta}}\right)^2\right]$, where $N_i$ is the number of time samples in pixel $i$.
The uncertainty on Juno's transmitter pointing direction is estimated to be around 0.25 degrees \citep{JunoPointingError}. Assuming a similar uncertainty for the microwave receivers, we estimate pointing errors from the above formula and a Juno-beam-convolved approximate sky model. Except for a few pixels with very few hits, the pointing errors are very small.
\end{enumerate}
 
By treating the map, offset, and gain solutions as a single vector, \cite{2010A&A...522A..94K} derives a covariance with terms (i) and (ii) from the standard map-making equation via blockwise matrix inversion. With the assumption of constant thermal noise, this map-space covariance is given by
\begin{linenomath*}\begin{equation}\label{eq:map_space_cov}
\left<\bm{m}^T\bm{m}\right> =   \sigma_N^2(\bm{P}^T\bm{P})^{-1} + (\bm{P}^T\bm{P})^{-1}(\bm{P}^T\bm{F})\left<\bm{a}^T\bm{a}\right>(\bm{P}^T\bm{F})^T(\bm{P}^T\bm{P})^{-1}.
\end{equation}\end{linenomath*} 
The second term describes the map-space covariance due to the uncertain noise and gain offsets. Sandwiched in the middle of this second term is the covariance of the noise and gain offset amplitudes, which is given by
\begin{linenomath*}\begin{equation}\label{eq:a_cov}
\left<\bm{a}^T\bm{a}\right> = \left[ \mathcal{P}^{-1}+ \frac{1}{\sigma_N^2}\bm{F}^T\bm{F} - \frac{1}{\sigma_N^2}(\bm{F}^T\bm{P})(\bm{P}^T\bm{P})^{-1}(\bm{F}^T\bm{P})^T \right]^{-1}.
\end{equation}\end{linenomath*}
In the above expression, $\mathcal{P}$ represents any assumed prior on the covariance of the noise and gain offsets. The only prior we assumed was that both the sum of the noise offsets and the sum of the gain offsets are constrained to be zero. This is included by placing large constant terms in $\mathcal{P}^{-1}$ on the upper left (noise offsets) and lower right (gain offsets) quadrants.  We note that, for the full gain solution map-maker, these covariance estimates use the penultimate iteration's map estimate, $\hat{\bm{m}}^{(n-1)}$, as a template for the true sky signal in the matrix $\bm{F}$. Although there will be errors in that map estimate, simulations confirm that, for the lower frequency bands, they are small compared to the magnitude of the sky, so that these formulas yield approximately correct gain fluctuation errors. For R3 through R6, the gain solution is too inaccurate to use, so for those bands the covariance model for $\left<\bm{a}^T\bm{a}\right>$ only follows equation \ref{eq:a_cov} for the noise offsets, while the gain offsets are assigned a standard deviation of $2\%$, consistent with both the expected level of the Juno gain errors and the MADAM gain solutions for R1 and R2. Based on the R1 and R2 solutions, the gain errors are assumed to be correlated over long time-scales, modeled as a Gaussian with a width of 50 days. When those gain errors are translated to the map covariance via equation \ref{eq:map_space_cov}, we use the PySM sky model (convolved with the Juno beam) to eliminate thermal noise in the sky template. We also subtract the mean of the PySM sky model over each scan to avoid double-counting offset errors (mean subtraction is not required for R1 and R2, because the full form of equation \ref{eq:a_cov} accounts for correlations between gain and noise offset solutions).  As written, equation \ref{eq:a_cov} assumes that there is no correlated noise between measurements at time-scales shorter than our 4-hour MADAM period. However, short time-scale drifts cannot be fully removed by the on-board calibration procedure, which leaves residual correlated noise at the $\sim$100 second time-scale over which the gain and offset are calibrated (see figure \ref{fig:noise_residual_auto_corr}). This correlated term likely dominates the uncertainty on gain and noise offset estimates from repeated measurements within the 4-hour MADAM period. In an attempt to approximately capture this effect, we rescale $\sigma_N^2$ in equation \ref{eq:a_cov} to match the angular power spectrum of errors in MADAM simulations at large angular scales (see section \ref{subsec:time-stream_sims}).

The full covariance matrix at $N_{side}=64$ is quite large and unwieldy. To compress it, we compute an SVD of the second term in equation \ref{eq:map_space_cov}. We find that approximately 50 terms account for 90 percent of its power. We release both the diagonal thermal noise term and this rank-50 SVD of the noise/gain offset term in map space. For both R1 and R2, we release an SVD of the combined gain/noise offset errors. For R3 through R6, we release separate SVDs for the noise offsets and the gain offsets.

\subsection{Summary of Systematic Errors}\label{subsection:systematic_error_list}
We estimate 5 sources of systematic errors.
\begin{enumerate}
\item \textbf{Monopole offset.} Pre-launch tests with a controllable blackbody calibration target at $\sim$300 Kelvin found that the zero-level of the radiometers could be calibrated to $\sim2\%$ or $\pm$6 Kelvin. Similar performance was confirmed after launch, where several Kelvin deviations from the expected $2.7$ Kelvin CMB blackbody temperature were observed at high frequencies \citep{janssenetal}. The Juno maps therefore must rely on external data sets to calibrate the map zero-levels. Details on the calibration choice are included in section \ref{subsec:abs_cal}.
\item \textbf{Gain offset.} The pre-launch tests with a controllable $\sim$300 Kelvin blackbody also provide an estimate of the expected gain offset, since the gain is calibrated with a noise diode of similar brightness temperature. The gain error is estimated to be $2\%$. 
\item \textbf{Pixelization errors.} Even after convolution by the large Juno beams, the maps for R1 and R2 at $N_{side}=64$ contain significant pixelization errors due to the steep angular gradient of the Galactic emission. These errors can be estimated by binning a high-resolution map template with the Juno scan strategy at $N_{side}=64$. They can also be greatly reduced by binning the data to maps at a higher angular resolution. We release simulated pixelization error maps for all bands and, for the 2 lowest bands, we also release maps at $N_{side}=512$, which reduces the pixelization errors to a level smaller than the thermal noise errors but also results in more un-mapped pixels.
\item \textbf{Polarization Leakage.} The single-polarized Juno receivers experience significant polarization leakage. To estimate its size and angular characteristics, simulated polarized contributions to the Juno maps are presented in section \ref{subsection:leakage_sim}.

\item \textbf{Spinning spacecraft smearing.} The 1-2 rpm spinning Juno scan changes the pointing direction of each antenna by about 1 degree over the 0.1 second integration time. This can be approximated as an effective convolution of the map in the ecliptic direction by a $\sim1$-degree tophat function, in addition to the regular beam convolution. 

\end{enumerate}

\section{Results and Error Model Verification}\label{section:results}
This section shows the final Juno maps, estimates of their errors, and tests to verify these error models.
\begin{figure*}
    \includegraphics[width=\textwidth]{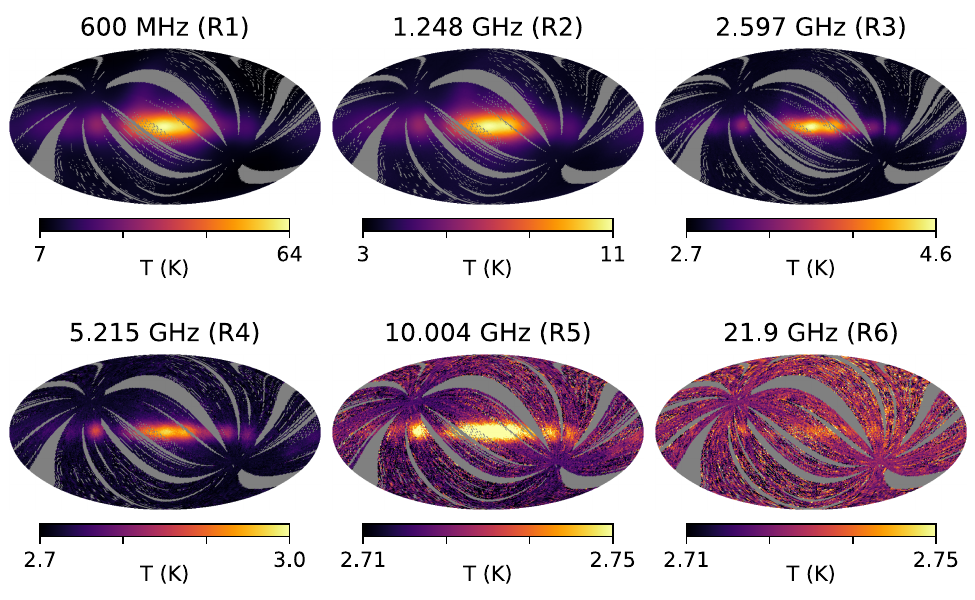}
    \caption{ \label{fig:Juno_maps_unconv} The 6 Juno maps, made at $N_{side}=64$. The 2 lowest frequency maps include MADAM noise offset and gain offset corrections. The highest 4 frequencies only include the MADAM noise offset de-striping. Pixels with fewer than 100 hits are removed. The monopole levels of the maps are calibrated according to the ARCADE 2 model of a radio background and CMB.}
\end{figure*}

\begin{figure*}
    \includegraphics[width=\textwidth]{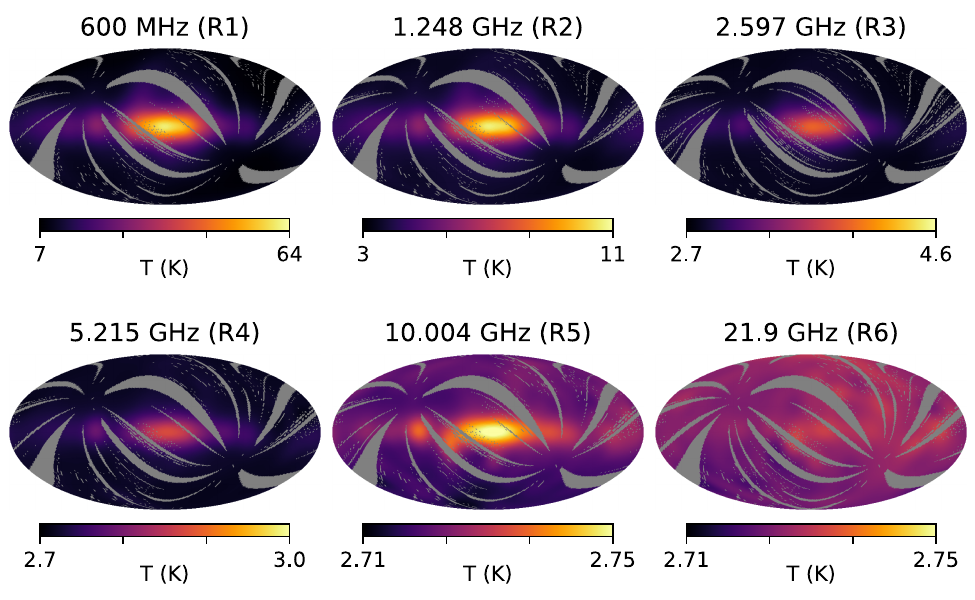}
    \caption{ \label{fig:Juno_maps_conv} The 6 Juno maps at $N_{side}=64$, convolved to a common Gaussian beam full-width-half-maximum of 21 degrees.}
\end{figure*}

\subsection{Map solutions}

The final map solutions are plotted in figure \ref{fig:Juno_maps_unconv}. Pixels with no coverage and the few high-noise pixels with fewer than 100 hits are greyed out. At this resolution, the R6 map appears to be noise-dominated, except for a slight increase in temperature in the Galactic plane. However, when all the maps are convolved to a common 21-degree FWHM Gaussian beam resolution (figure \ref{fig:Juno_maps_conv}), the CMB dipole is visible in both R5 and R6. To deal with missing coverage before computing the beam convolution, each empty map pixel is filled in with the average of present pixels in a 10-degree radius circle. After convolution, all pixels with zero sky coverage are re-masked.

\subsection{Absolute Calibration Procedure}\label{subsec:abs_cal}

The zero-level of the maps is calibrated to match that fit by the ARCADE 2 team \citep{2011ApJ...734....5F} from partial sky measurements in bands from $3.2\,GHz$ to $90\,GHz$, from FIRAS measurements of the CMB at $250\,GHz$, and from low-frequency all-sky radio maps. They model the sky as a Milky Way template plus a $2.725\,K$ CMB and a $24.1(\nu/310\,MHz)^{-2.599}\,K$ radio background. The resulting monopole level presented in the Juno maps is simply a calibration choice and should not be viewed as evidence for or against the existence of a radio background.

No gain correction factor is applied since several Juno maps are in frequency bands that have not been observed across the full sky before. The expected absolute accuracy of the map fluctuation level is $2\%$ \citep{janssenetal}.

\subsection{Residual Noise Analysis}\label{subsec:residual_noise_analysis}

\begin{figure}
    \includegraphics[width=\columnwidth]{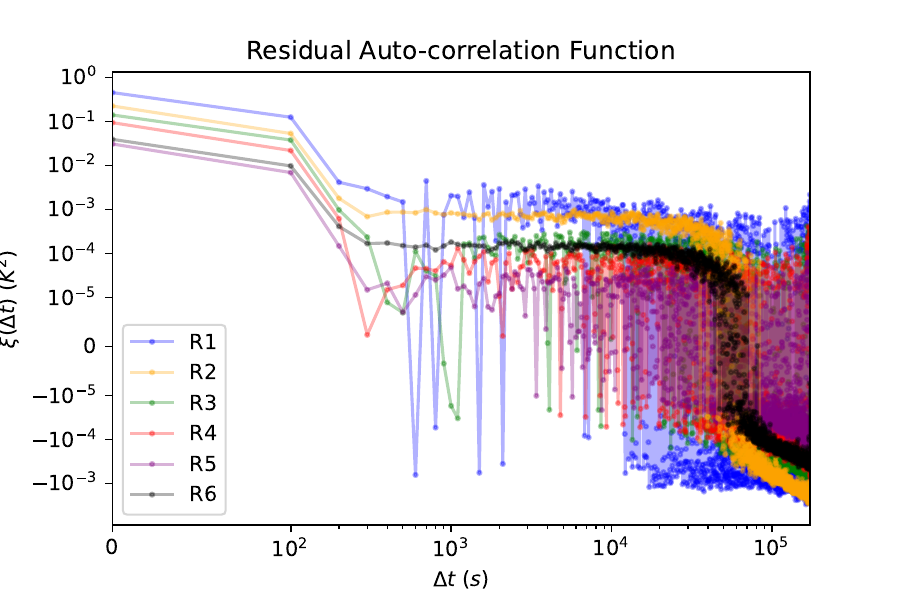}
    \caption{\label{fig:noise_residual_auto_corr} The auto-correlation function of the time-stream noise residuals for all Juno bands. The y-axis varies linearly between $\pm10^{-5}\,K^2$ and logarithmically outside those bounds. The time axis varies linearly between $0$ and $100$ seconds and logarithmically elsewhere. The level at zero time difference is due mainly to the diagonal thermal noise, roughly consistent with the level predicted in table \ref{table1}. There is also a smaller short time-scale correlated component, visible in all bands at 100 seconds, which is due to the on-board gain and noise offset calibration procedure described in section \ref{subsection:on_board_cal}. This correlated component also contributes some of the variance at zero time lag. Noise residuals at time-scales between 100 seconds and 4 hours are generally positively correlated, with a magnitude roughly 100 times smaller than the calibration noise level. This is consistent with the expected noise level on the MADAM noise/gain amplitudes, assuming that the calibration noise dominates those measurements. For most channels, the positive correlations cease at time-scales longer than the 4-hour scale of the MADAM periods.}
\end{figure}

\begin{figure*}
    \includegraphics[width=\textwidth]{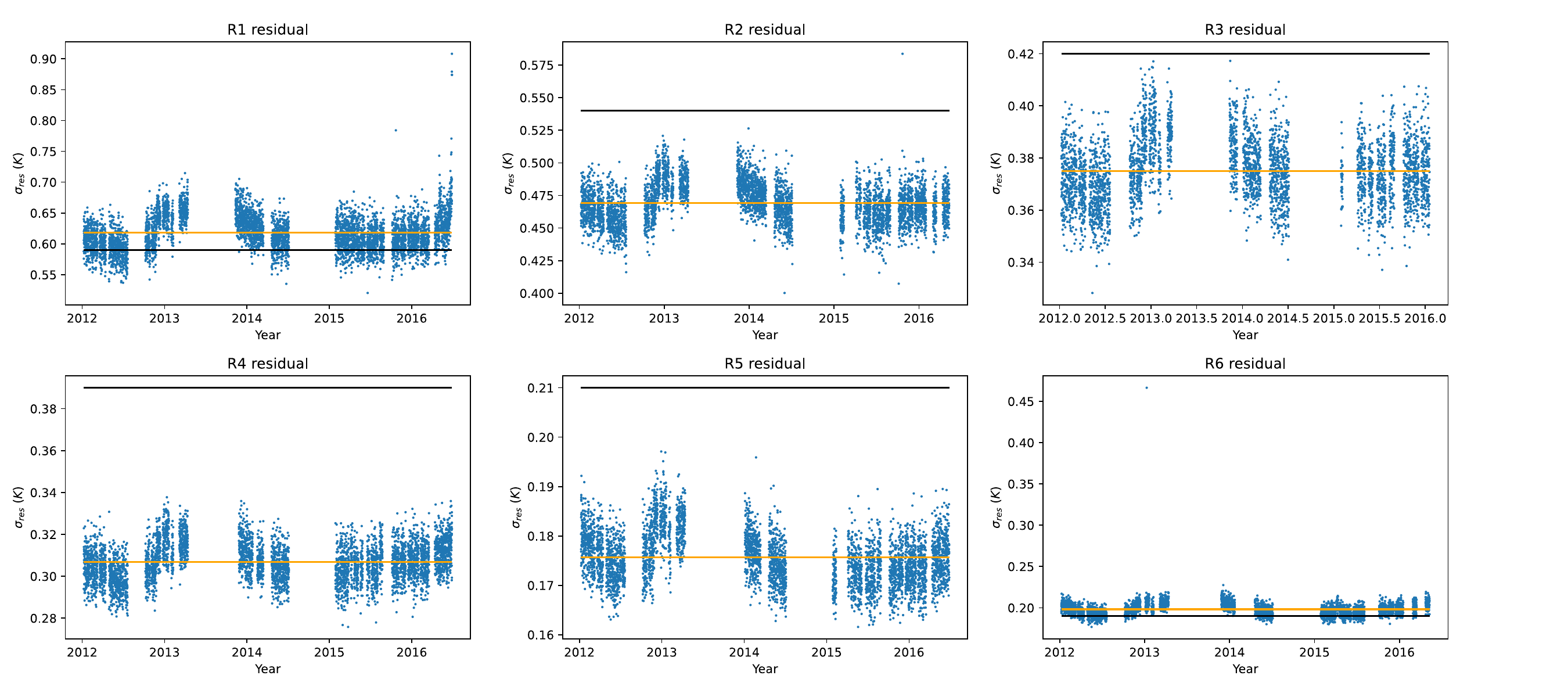}
    \caption{ \label{fig:residual_variance} A scatter plot of the standard deviation of the time-stream residual for each band and every 4-hour period of the dataset. The black horizontal line shows the expected thermal noise level from table \ref{table1}. The orange horizontal line is a fit to the noise level from the residual. The fit mainly reflects the level of diagonal thermal noise, but it also includes small levels of added variance from short-time-scale calibration noise, pointing errors, long-time-scale gain and noise solution uncertainties, and pixelization noise. To ensure that pixelization noise is small, the R1 and R2 bands use the $N_{side}=512$ map solution to compute the residual. The other bands, where a dimmer Galactic plane results in smaller pixelization noise, use the $N_{side}=64$ maps. The scatter in the variance computed from scan to scan is slightly larger than one would expect from uncorrelated Gaussian noise in the approximate 10,000 samples per scan due to the the on-board calibration procedure, which has roughly 100 times fewer independent measurements per scan. The long time-scale drifts in the envelope of residual standard deviations, which appear to be correlated between bands, are not well understood. Generally, however, the noise levels of the instruments are as expected.}
\end{figure*}

Once map solutions are found, the time-stream residual is computed by subtracting the map solution, noise offset solution, and gain solution from the time-stream data. 
\begin{linenomath*}\begin{equation}
\bm{y}_{res} = \bm{y} - \bm{P}\hat{\bm{m}} - \bm{F}\hat{\bm{a}}.
\end{equation}\end{linenomath*}
To test for residual correlated noise, we compute the auto-correlation function of the time-stream residual, $\xi(\Delta t)$, on time-scales from a few minutes to several days. Figure \ref{fig:noise_residual_auto_corr} shows this auto-correlation for each band. The auto-correlation is dominated by a spike at zero time lag, due mainly to diagonal thermal noise. In addition, correlated noise at the shortest non-zero binned time-scale, which is 100 seconds, is found at roughly a quarter of the magnitude of the zero-lag auto-correlation. This short time-scale correlated noise is likely due to the on-board calibration procedure described in section \ref{subsection:on_board_cal}, where short time-scale fluctuations in the gain, system temperature, and DC offset are removed. In this procedure, both the diode and internal load are measured once every 10 seconds, but these are averaged over time-scales of a few minutes. For time-scales longer than 100 seconds but shorter than 4 hours, the noise auto-correlation for all bands tends towards a constant positive value of $\sim1/100$ the level of the calibration noise. This is the expected uncertainty on the 4-hour gain/offset amplitudes computed by the MADAM map-making procedure, which is limited by the $\sim100$ independent calibrations within each 4-hour period. At time-scales longer than 4 hours, for most bands, the auto-correlation fluctuates between positive and negative values, as expected for time-scales greater than the MADAM period length. The shift from positive to negative correlations near 20 hours for R2 and R6 is not understood.

We also compute the standard deviation of the time-stream residual for each 4-hour period of the scan. Figure \ref{fig:residual_variance} shows a scatter plot of this standard deviation for each band. The noise levels are consistent with the expected thermal noise (table \ref{table1}).

\begin{figure*}
    \includegraphics[width=\textwidth]{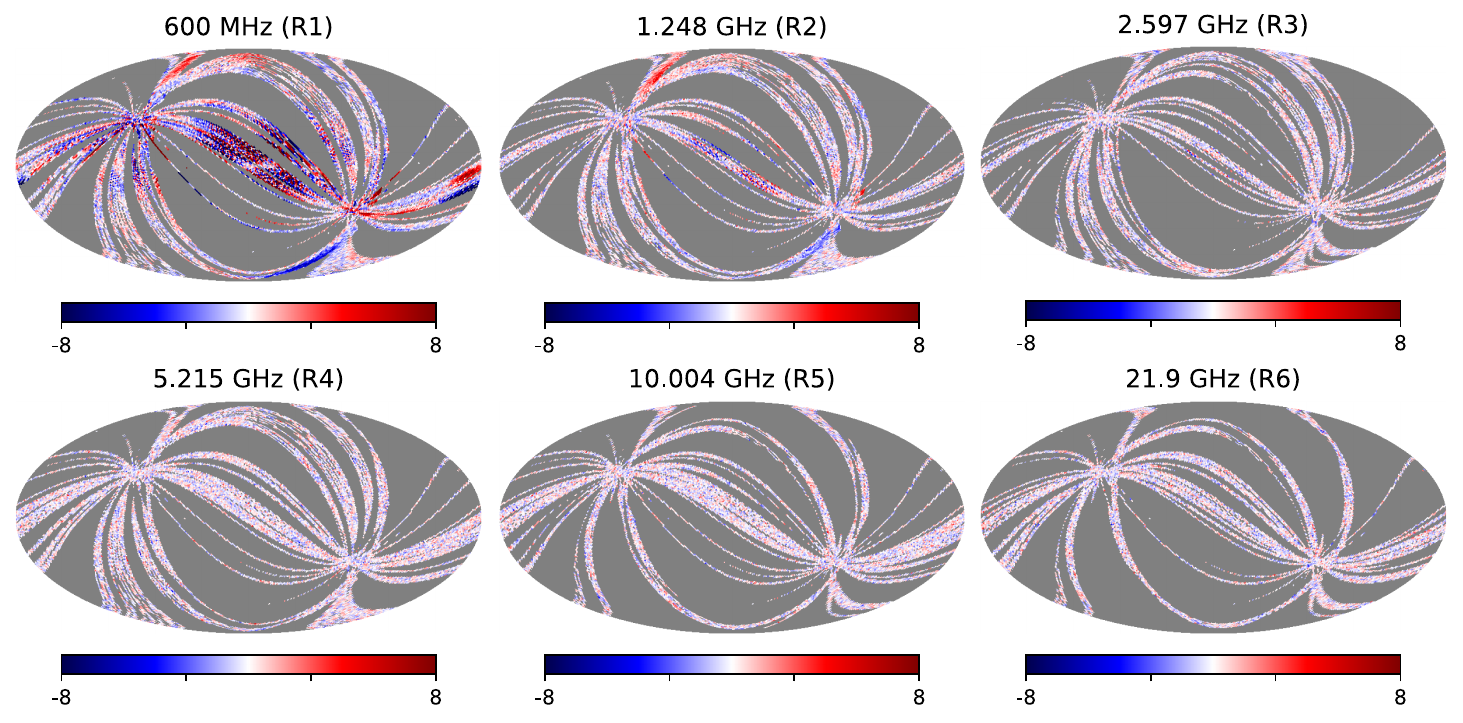}
    \caption{\label{fig:real_whitened_map_differences} Differences between maps made with half of the data, whitened by dividing the differences by the square root of a simple diagonal noise model. The diagonal noise model comes from the hitmap and the time-stream residual variance. The scale of the plots is saturated at the $\pm8$-$\sigma$ level. Maps R3 through R6 are consistent with white-noise simulations (compare these difference maps with the white noise simulations shown in figure \ref{fig:sim_white_map_differences}). Maps R1 and R2 show deviations from the diagonal noise model, driven by pixelization noise, which is strongest in the plane of the Galaxy (see figures \ref{fig:pix_errors} and \ref{fig:R1_pix_err_diff}).  }
\end{figure*}

\begin{figure*}
    \includegraphics[width=\textwidth]{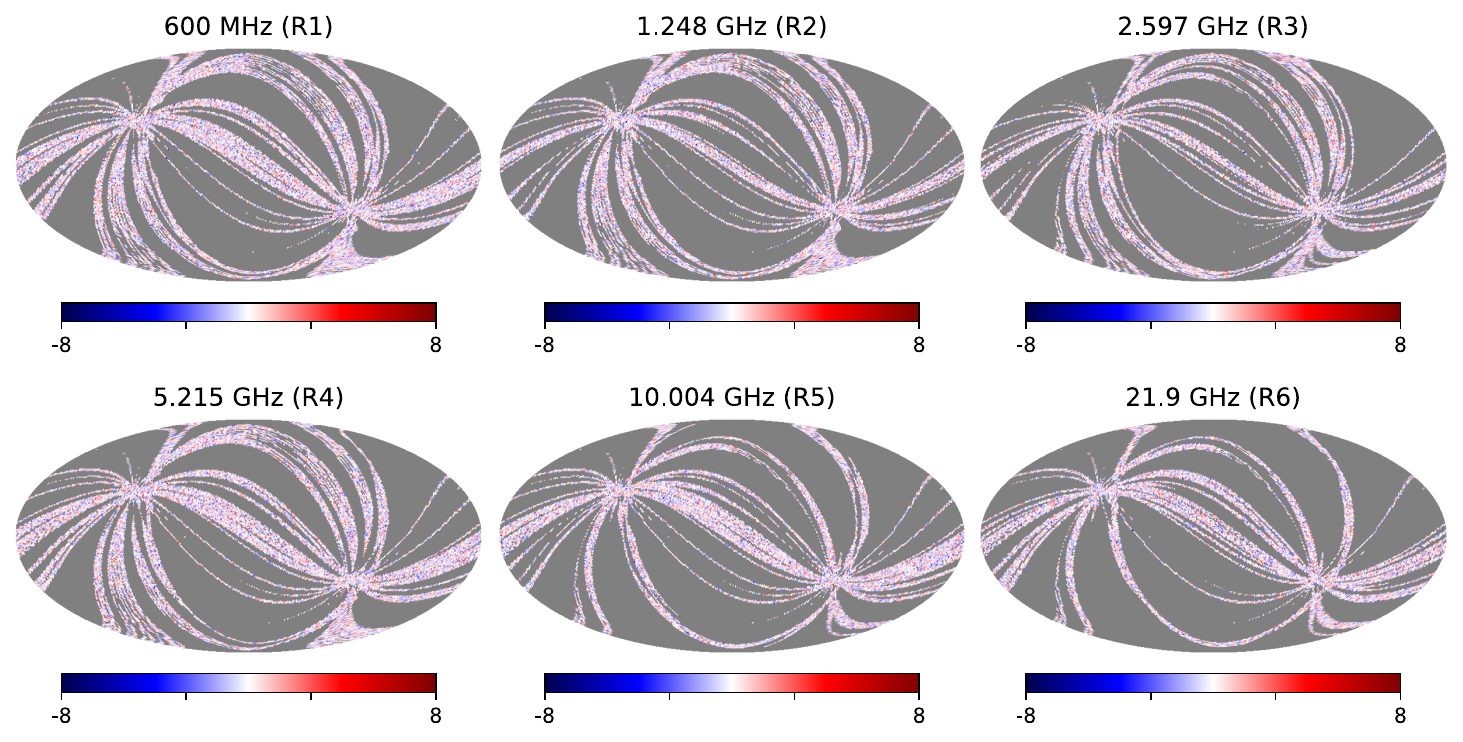}
    \caption{\label{fig:sim_white_map_differences} Simulated whitened differences between maps made with half of the data, where the simulated differences are purely due to diagonal thermal noise. As in figure \ref{fig:real_whitened_map_differences}, the scale of the plots is saturated at the $\pm8$-$\sigma$ level.  }
\end{figure*}

\begin{figure*}
    \includegraphics[width=\textwidth]{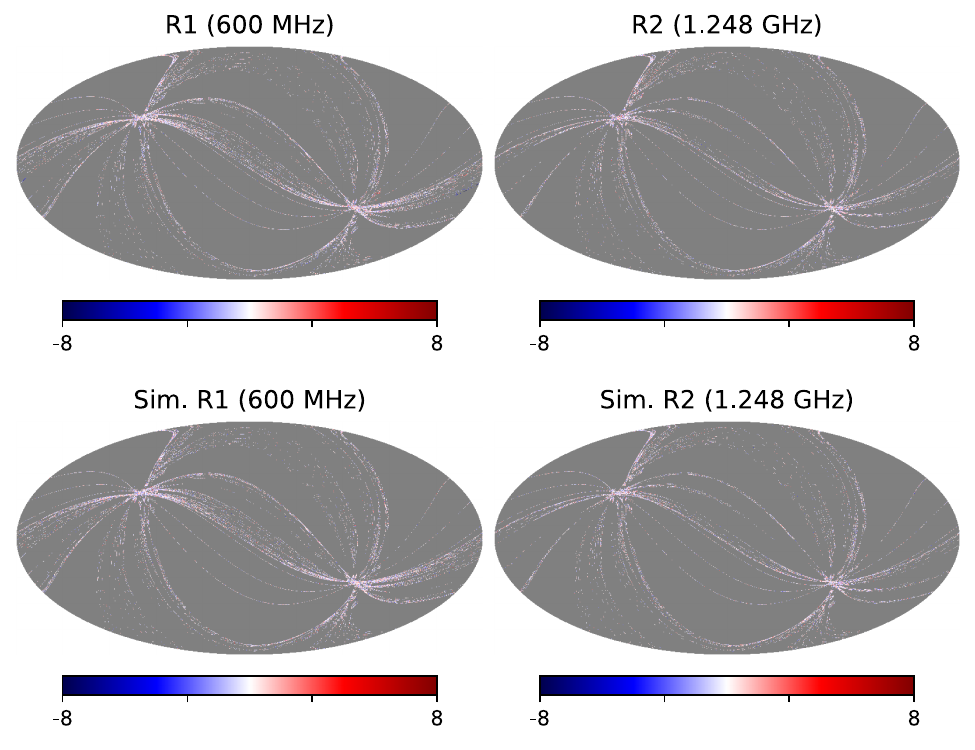}
    \caption{\label{fig:high_res_map_differences} 
    Top: real whitened differences between maps made with half of the data for R1 and R2 at $N_{side}=512$, whitened by dividing the differences by the square root of a simple diagonal noise model. Bottom: simulated whitened differences between $N_{side}=512$ R1 and R2 maps made with half of the data, where the simulated differences are purely due to diagonal thermal noise. The scale of the plots is saturated at the $\pm8$-$\sigma$ level. The strong pixelization errors that were present at $N_{side}=64$ for these 2 bands have subsided below the thermal noise at this resolution, and the real map differences appear consistent with the simulation.}
\end{figure*}

Lastly, we compute map-space residuals by subtracting maps made with different periods of the time-stream data, but with a common MADAM solution for the gain and noise offsets. Figure \ref{fig:real_whitened_map_differences} shows map differences divided by a diagonal noise model consistent with the hit map and the average variance of the time-stream residuals. Figure \ref{fig:sim_white_map_differences} shows simulated versions of these map differences for that simple diagonal thermal noise model. Map differences for R3 through R6 are well described by the diagonal noise model, but map R2 and especially map R1 deviate from this model. The deviation is driven by pixelization noise of the bright Galactic plane in those bands. Difference maps of those two bands at the higher $N_{side}=512$ resolution, shown in figure \ref{fig:high_res_map_differences}, have reduced the pixelization errors to below the thermal noise level, and they show errors consistent with a diagonal thermal noise model. Table \ref{table2} shows the $\chi^2$ per pixel of the map differences using only this simple thermal noise model. The values support the interpretation that, at a sufficiently high resolution, the map differences are well described by diagonal thermal noise. However, because they use a common MADAM solution, these map differences are missing some correlated noise, whose amplitude we characterize via our full time-stream simulations.

\begin{table}\label{table2}
\tabcolsep=0.11cm
\setlength\tabcolsep{8pt}
\begin{tabular}{@{\extracolsep{\fill}} c c c }
\toprule
 Receiver  & $\chi^2$ per pixel, $N_{side}=64$   & 
  $\chi^2$ per pixel, $N_{side}=512$ \\ 
 \midrule
 R1 & 12.69 & 0.99 \\  
 R2 & 1.51 & 0.99 \\ 
 R3 & 1.04 & - \\ 
 R4 & 0.99 & - \\ 
 R5 & 0.99 & - \\ 
 R6 & 1.01 & - \\ 
 \bottomrule 
\end{tabular}
\caption{Summary of the $\chi^2$ per pixel of Juno difference maps, made by subtracting maps made with half the data from maps made with the other half. The noise model used to compute $\chi^2$ is a simple diagonal noise model using the Juno hitmaps and the magnitude of the time-stream residual variance. This model is a good description of the errors at $N_{side}=64$ for difference maps R3 through R6. For R1 and R2, pixelization errors are non-negligible at $N_{side}=64$. They become sub-dominant to thermal noise at $N_{side}=512$. At that pixel scale, the simple thermal noise model is a good description for R1 and R2.}
\end{table}

\subsection{Time-stream simulations}\label{subsec:time-stream_sims}
We validate our implementation of the MADAM map-maker and our map-space noise model by analyzing synthetic maps made from simulated time-stream data. From the residual noise analysis of the previous section, the Juno time-stream data is well described by diagonal thermal noise, correlated calibration noise on a 100-second time-scale, and noise/gain drifts on time-scales longer than 4 hours. Juno data is therefore simulated as consisting of the following components:
\begin{itemize}
    \item Thermal noise, with a variance equal to the zero-lag value of the time-stream residual autocorrelation minus the value of the 100-second autocorrelation. 
    \item Calibration noise, fully correlated at the 100-second time-scale, with a variance equal to the measured value of the 100-second time-stream residual autocorrelation. 
    \item Slow 4-hour noise drifts.
    \item The Python Sky Model (PySM) \citep{pysm} multiplied by $1 + \delta g$, where $\delta g$ are slow 4-hour drifts in the gain. 
\end{itemize}
Values for the input gain and noise drifts in all the bands come from the actual solved gain and noise drifts from R1 where, as these simulations show, the signal-to-noise ratio is sufficient to solve for both. 

\begin{figure*}
    \includegraphics[width=\textwidth]{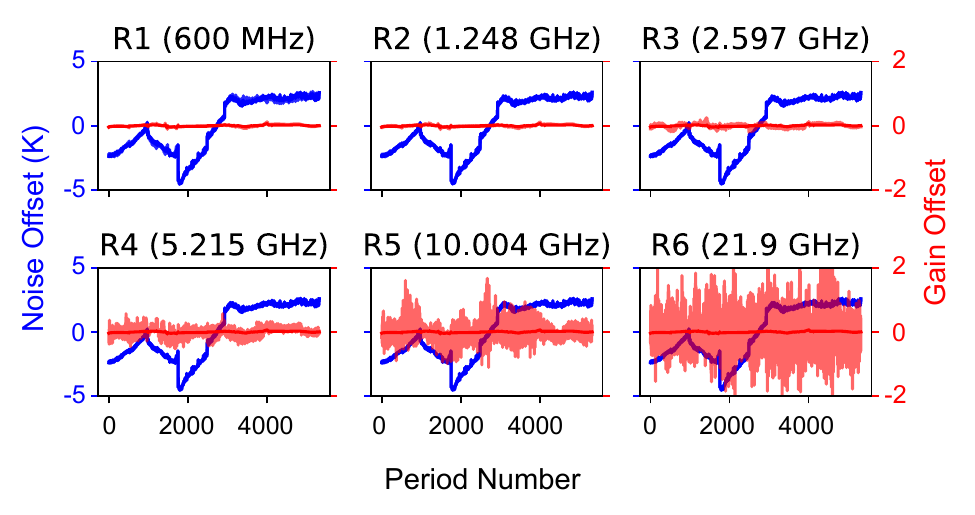}
    \caption{ \label{fig:noise_and_gain_sols_sim} Simulated MADAM-computed noise and gain offset solutions. Simulations included the expected thermal noise level and PySM model maps of the Galaxy for each band. The dark blue and red curves show the input noise and gain offsets to the simulation. The transparent blue and red curves show the noise and gain offset solutions found by MADAM. All six bands find accurate solutions to the noise offset amplitudes. For the gain offset amplitudes, the noise in the solution grows with increasing frequency due to the decreasing brightness of the Galactic plane relative to the thermal noise level.}
\end{figure*}

The solved MADAM noise and gain offset amplitudes for the simulation are plotted in figure \ref{fig:noise_and_gain_sols_sim}. All bands find the correct noise offsets. The recovered gain drift solutions become noisy for bands higher than R2, due to the decreasing brightness of the Galactic plane with higher frequency. These simulations informed our choice to forego using gain drift solutions in bands R3 through R6.

\begin{figure*}
    \includegraphics[width=\textwidth]{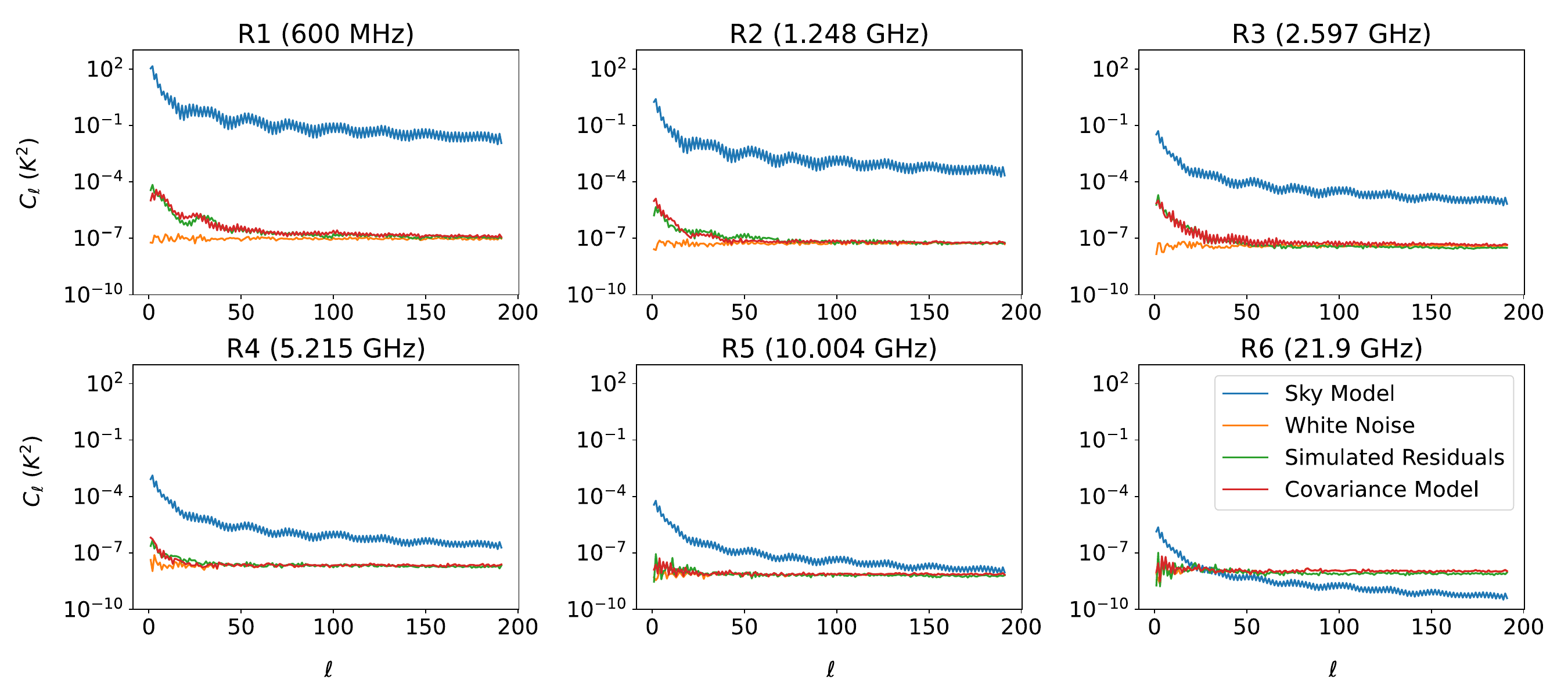}
    \caption{ \label{fig:noise_pcl} The weighted angular power spectrum of the sky, white noise model, simulated noise residuals, and our map-space covariance model for each of the six Juno bands. Before computing the angular power spectrum, maps are weighted by radiometric inverse variance weights, normalized so that the average value of the weights is one. No un-mixing of these weighted power spectra is performed.}
\end{figure*}

We compute the map-space residual of our simulations by subtracting the full simulated map solution for each band from a noiseless simulation that followed the same Juno scan pattern. This residual provides a realization of the map-space statistical noise without any pixelization noise. The residual is then used to set the scale of the correlated map-space error model (the second term in equation \ref{eq:map_space_cov}). The scale is adjusted so that the weighted angular power spectrum of random draws from our SVD-compressed covariance model approximately matches that of the simulated map-space residual. Figure \ref{fig:noise_pcl} shows the angular power spectra of this noise model compared to our simulations, along with the power spectrum of the PySM sky model.

\subsection{Pixelization Noise Corrections}\label{subsection:pixelization_noise}
\begin{figure*}
    \includegraphics[width=\textwidth]{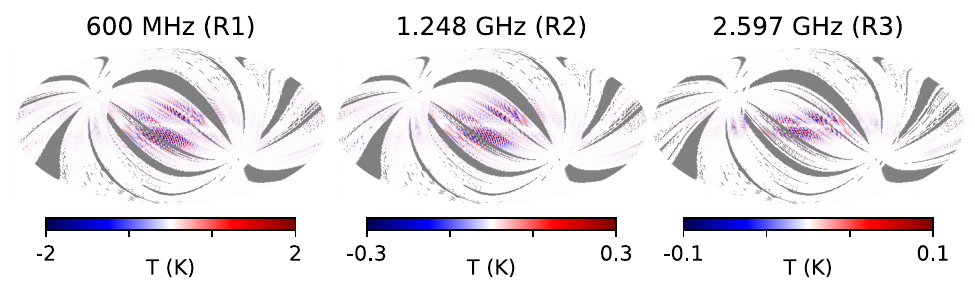}
    \caption{\label{fig:pix_errors} Simulated pixelization noise for bands R1 through R3, for maps made at a resolution of $N_{side}=64$. This correction can be applied directly to the maps to approximately remove the pixelization error. Pixelization corrections are provided for all bands, but the plots for R4 through R6 are omitted for brevity. Pixelization errors are below the level of thermal noise for R3 through R6.}
\end{figure*}

\begin{figure*}
    \includegraphics[width=\textwidth]{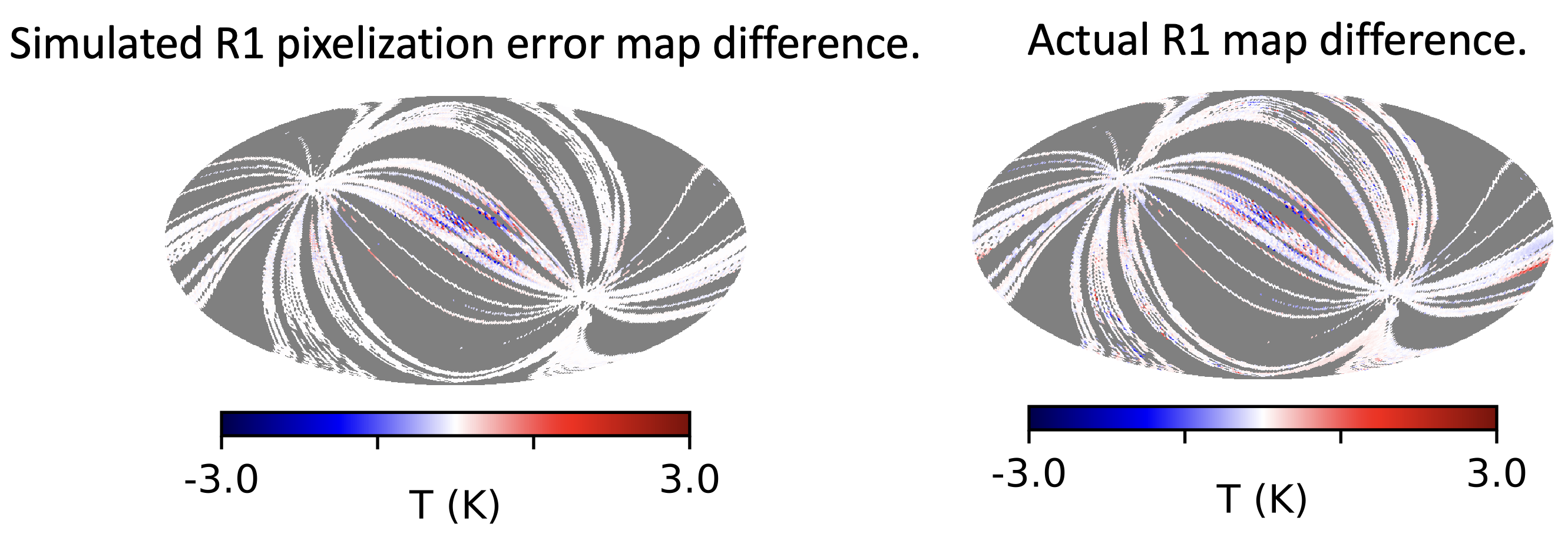}
    \caption{\label{fig:R1_pix_err_diff} Right: Difference between maps of R1 (600 MHz) made with half the data. Left: Simulated difference between R1 (600 MHz) maps made from binning PySM maps according to Juno's scan for half the data. In this simulation, no thermal noise is included, only the effect of map pixelization. Maps shown are made at a resolution of $N_{side}=64$.}
\end{figure*}

There will inevitably be some error in the maps due to the finite pixel size and the effect of the pointing matrix to average non-uniformly spaced telescope pointings within each pixel \citep{mapcumba, poutanen2006comparison}. Given a high-resolution sky model, a simple formula for the pixelization error in the time-stream data is derived in \cite{mapcumba}. We compute this same pixelization error in map-space, by down-grading a high-resolution ($N_{side}=2048$) Juno-beam-convolved PySM map to $N_{side}=64$ to serve as a `perfect' low-resolution map, for which sampling within each pixel is homogeneous. We also form an estimate of the sky as seen by Juno with pixelization errors, by applying the actual Juno pointing information to bin the same Juno-beam-convolved high-resolution template to a map at $N_{side}=64$. Subtracting the Juno-binned map from the `perfect'  map template yields an estimate of the correction one can add to the Juno maps to approximately remove the pixelization errors. Since this correction relies on an external model of the sky, we have not applied this correction to any of the released Juno maps. We instead release it as a separate data product that can be added to the maps if desired. 
This correction factor is plotted in figure \ref{fig:pix_errors} for R1 through R3. The pixelization errors can also be computed for any subset of the Juno scan data. By comparing simulated pixelization error differences to map differences made with actual subsets of data, we confirm that pixelization errors dominate the map differences near the Galactic plane for bands R1 and R2. Figure \ref{fig:R1_pix_err_diff} shows this effect for R1. For R3 through R6, the Galactic plane is dim enough compared to the thermal noise in the maps that pixelization errors are sub-dominant.

\subsection{Simulated Polarization Leakage}\label{subsection:leakage_sim}

\begin{figure*}
    \includegraphics[width=\textwidth]{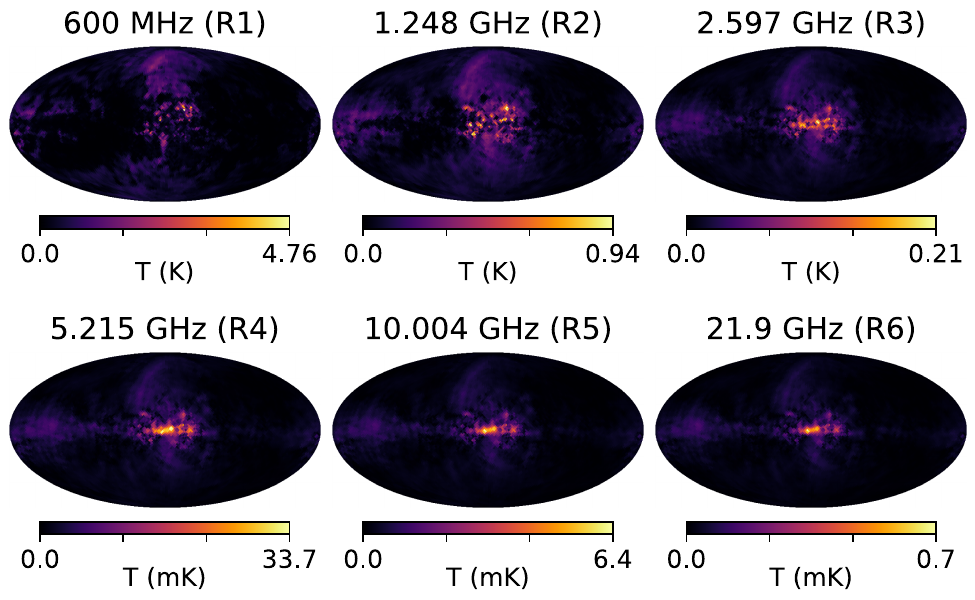}
    \caption{\label{fig:pysm_pol_sim} The magnitude of polarized emission, $\sqrt{Q^2+U^2}$, for this polarized sky model. The model is based on PySM, which uses polarized synchrotron and dust templates from WMAP 23 GHz observations. We also simulate the effect of Faraday rotation and bandwidth de-polarization, using the Milky Way Faraday depth map of \cite{2022A&A...657A..43H}. Polarized Milky Way emission is assumed to be evenly distributed along Faraday depths from zero out to the total depth measured in the Milky Way map. These maps are made at $N_{side}=32$, which suppresses some small-scale structure. However, this resolution is sufficient for simulating Juno's response, since the Juno leakage patterns are on the 10 to 20 degree scale.}
\end{figure*}

\begin{figure*}
    \includegraphics[width=\textwidth]{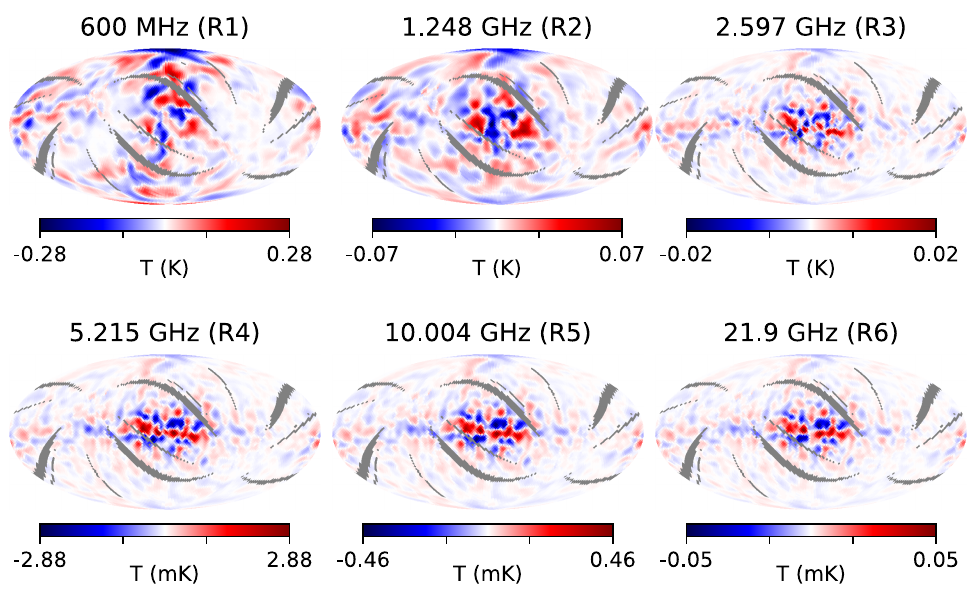}
    \caption{\label{fig:pysm_pol_sim_conv}
    Calculated contribution of leakage from the simulated polarized Stokes $Q$ and $U$ maps to the Stokes $I$ maps for the 6 Juno bands. The leakage is computed by convolving the average pointed and rotated Juno leakage patterns (see figure \ref{fig:polarized_beams}) with simulated $Q$ and $U$ maps, whose summed RMS magnitude is shown in figure \ref{fig:pysm_pol_sim}. The large Juno leakage beam patterns greatly reduce the magnitude of polarization leakage present in the final $I$ maps, via beam de-polarization. At all frequencies, the polarization leakage is well below the total magnitude of the Juno $I$ maps.}
\end{figure*}

Since the Juno receivers are all singly polarized, there is potential for significant polarization leakage to contaminate the maps. To estimate the magnitude of this leakage, we simulate Juno's response to a model of the polarized sky. Since there is considerable uncertainty in the true polarized sky, especially at the lower wavelengths, these leakage response maps should be viewed as only an approximation to the magnitude and shape of the leakage contamination.

The chosen model for the polarized sky comes from the Python Sky Model (PySM) \citep{pysm}. We use the s3 and d3 models of synchrotron and dust emission. However, these polarized emission templates are based on WMAP observations at 23 GHz, a band in which the effect of Faraday rotation is small. Polarized observations from the Parkes radio telescope at 2.3 GHz \citep{10.1093/mnras/stz806} show substantially reduced polarized emission near the Galactic plane, due to the high Faraday depth of the Galactic plane and bandwidth depolarization of Faraday rotated polarized emission. In an attempt to model this effect, we use the Milky Way Faraday depth map, based on observations of extragalactic sources, from \cite{2022A&A...657A..43H}. We assume that Milky Way emission is uniformly produced over all Faraday depths from 0 to the measured total Faraday depth from extragalactic sources. We then integrate the Faraday rotated PySM templates over the Juno bands. The resulting polarization maps are shown in figure \ref{fig:pysm_pol_sim}. Faraday rotation and bandwidth de-polarization are significant for R1 through R3, especially near the Galactic plane, where the Faraday depth is high. Due to the shorter wavelengths of R4 through R6, Faraday rotation in those bands is insignificant, and the Faraday rotated maps are nearly the same as the un-rotated PySM templates.

We simulate Juno's response to polarized emission by convolving the Faraday rotated PySM maps by the rotated Juno beam leakage patterns in real space. This convolution can be computed quickly because the polarized leakage patterns (see figure \ref{fig:polarized_beams}) are well approximated as simple separable functions of $\theta$ and $\phi$. The $\theta$-dependence is just the 2D Gaussian function with FWHM given by table \ref{table1}. The $\phi$-dependence is $\cos(2\phi - 2\psi)$ for $M_{IQ}$ and $-\sin(2\phi - 2\psi)$ for $M_{IU}$, where $\psi$ is the receiver rotation angle from the Galactic North direction, $-\hat{\theta}$. Using the rotation angle data for the full Juno data set along with sine and cosine trigonometric identities, a scan averaged $\phi$-dependence for $M_{IQ}$ and $M_{IU}$ is computed for each pixel of the sky. This averaged leakage beam is moved to the correct sky pixel, using the healpy package, and the overlap with the PySM $Q$ and $U$ maps is computed. This is done for each pixel until the leakage map is complete. The simulated contribution of polarization leakage to the Juno maps is shown in figure \ref{fig:pysm_pol_sim_conv}.

These results show that polarization leakage is well below the total magnitude of the $I$ maps at all frequencies.  This relatively low leakage level is achieved via beam depolarization in all bands and bandwidth depolarization of Faraday rotated emission in the lower frequency bands.

\section{Discussion and Conclusion}\label{section:discussion}

We release 6 all-sky maps, and an estimate of their associated errors, with the hope that they can be used by the broader astrophysical community. Previous Juno publications, in which some all-sky maps appear, focused on Jupiter planetary science \citep{janssenetal}. Those sky maps were made primarily for calibration purposes, and the published images include interpolation to fill in gaps and smoothing to reduce noise. The maps presented in this paper differ in the details of the de-striper algorithm used, the removal of long time-scale gain drifts for maps R1 (600 MHz) and R2 (1.248 GHz), and the production of statistical and systematic error estimates. Although these Juno maps are limited in angular resolution, because they are made from a single instrument that rapidly surveyed great circles on the sky, their large-scale structure is robust and free of significant systematic uncertainties. 

The 600 MHz map provides a check for the large-scale structure in the well-used 408 MHz Haslam map \citep{haslam1982408}, which was stitched together from four separate maps. All of the Juno maps will provide measurements of the angular and frequency variation of the synchrotron spectral index, over a range where the spectral index is steepening with increasing frequency \citep{kogut2012synchrotron}. The four maps from 2.6 GHz to 21.9 GHz may be especially useful for component models of Galactic emission, which suffer from a near order-of-magnitude gap in frequency coverage between $2.3$ GHz $22.8$ GHz \citep{de2008model, zheng2017improved, pysm}. The three highest frequency bands may be useful for constraining the spectrum of AME emission.  For instance, large-scale fits to distinguish AME emission from other components in a Planck analysis \citep{adam2016planck} was hindered by lack of frequency coverage from 5-20 GHz \citep{dickinson2018state}.

These maps demonstrate the robustness of the MADAM/NPIPE map-making technique for removing long time-scale additive and multiplicative drifts. The significant pixelization errors found in the low-frequency maps highlight the need for careful analysis when characterizing continuum foregrounds. For example, when combining data sets of different frequencies made with different instruments, the different pixelization errors of the two scans could easily be misinterpreted as angular dependence of the synchrotron spectral index.

\section{Data Availability}\label{section:data_availability}
The maps and their associated covariances, pixelization error estimates, and polarization contribution estimates are publicly available on NASA's LAMBDA archive at \url{https://lambda.gsfc.nasa.gov/product/foreground/fg_juno_get.html}.

\section*{Acknowledgements}
 We thank the Juno PI Scott Bolton for their support of this work. We thank the reviewer for questions that led to a better understanding of the noise model. C. Anderson, P. Berger, and T.-C. Chang acknowledge support by NASA ROSES grants 17-ADAP17-0234 and 21-ADAP21-0122. Part of this work was done at the Jet Propulsion Laboratory, California Institute of Technology, under a contract with the National Aeronautics and Space Administration. This work utilized the software packages astropy \citep{astropy1, astropy2}, HEALPix \citep{healpix}, and healpy \citep{healpy}. \textcopyright $\,$2024. California Institute of Technology. Government sponsorship acknowledged.

\bibliography{main}{}

\begin{thebibliography}{}
\makeatletter
\relax
\def\mn@urlcharsother{\let\do\@makeother \do\$\do\&\do\#\do\^\do\_\do\%\do\~}
\def\mn@doi{\begingroup\mn@urlcharsother \@ifnextchar [ {\mn@doi@}
  {\mn@doi@[]}}
\def\mn@doi@[#1]#2{\def\@tempa{#1}\ifx\@tempa\@empty \href
  {http://dx.doi.org/#2} {doi:#2}\else \href {http://dx.doi.org/#2} {#1}\fi
  \endgroup}
\def\mn@eprint#1#2{\mn@eprint@#1:#2::\@nil}
\def\mn@eprint@arXiv#1{\href {http://arxiv.org/abs/#1} {{\tt arXiv:#1}}}
\def\mn@eprint@dblp#1{\href {http://dblp.uni-trier.de/rec/bibtex/#1.xml}
  {dblp:#1}}
\def\mn@eprint@#1:#2:#3:#4\@nil{\def\@tempa {#1}\def\@tempb {#2}\def\@tempc
  {#3}\ifx \@tempc \@empty \let \@tempc \@tempb \let \@tempb \@tempa \fi \ifx
  \@tempb \@empty \def\@tempb {arXiv}\fi \@ifundefined
  {mn@eprint@\@tempb}{\@tempb:\@tempc}{\expandafter \expandafter \csname
  mn@eprint@\@tempb\endcsname \expandafter{\@tempc}}}

\bibitem[\protect\citeauthoryear{Adam et~al.,}{Adam
  et~al.}{2016}]{adam2016planck}
Adam R.,  et~al., 2016, Astronomy \& Astrophysics, 594, A10

\bibitem[\protect\citeauthoryear{Akrami et~al.,}{Akrami
  et~al.}{2020a}]{akrami2020planck}
Akrami Y.,  et~al., 2020a, Astronomy \& Astrophysics, 641, A4

\bibitem[\protect\citeauthoryear{Akrami et~al.,}{Akrami et~al.}{2020b}]{NPIPE}
Akrami Y.,  et~al., 2020b, Astronomy \& Astrophysics, 643, A42

\bibitem[\protect\citeauthoryear{{Astropy Collaboration} et~al.,}{{Astropy
  Collaboration} et~al.}{2013}]{astropy1}
{Astropy Collaboration} et~al., 2013, \mn@doi [\aap]
  {10.1051/0004-6361/201322068}, \href
  {https://ui.adsabs.harvard.edu/abs/2013A&A...558A..33A} {558, A33}

\bibitem[\protect\citeauthoryear{{Astropy Collaboration} et~al.,}{{Astropy
  Collaboration} et~al.}{2018}]{astropy2}
{Astropy Collaboration} et~al., 2018, \mn@doi [\aj] {10.3847/1538-3881/aabc4f},
  \href {https://ui.adsabs.harvard.edu/abs/2018AJ....156..123A} {156, 123}

\bibitem[\protect\citeauthoryear{Bernardi et~al.,}{Bernardi
  et~al.}{2009}]{bernardi2009foregrounds}
Bernardi G.,  et~al., 2009, Astronomy \& Astrophysics, 500, 965

\bibitem[\protect\citeauthoryear{Carretti et~al.,}{Carretti
  et~al.}{2019}]{10.1093/mnras/stz806}
Carretti E.,  et~al., 2019, \mn@doi [Monthly Notices of the Royal Astronomical
  Society] {10.1093/mnras/stz806}, 489, 2330

\bibitem[\protect\citeauthoryear{Condon, Cotton, Greisen, Yin, Perley, Taylor
  \& Broderick}{Condon et~al.}{1998}]{nvss}
Condon J.~J.,  Cotton W.,  Greisen E.,  Yin Q.,  Perley R.~A.,  Taylor G.,
  Broderick J.,  1998, The Astronomical Journal, 115, 1693

\bibitem[\protect\citeauthoryear{Dickinson et~al.,}{Dickinson
  et~al.}{2018}]{dickinson2018state}
Dickinson C.,  et~al., 2018, New Astronomy Reviews, 80, 1

\bibitem[\protect\citeauthoryear{{Dor{\'e}}, {Teyssier}, {Bouchet}, {Vibert}
  \& {Prunet}}{{Dor{\'e}} et~al.}{2001}]{mapcumba}
{Dor{\'e}} O.,  {Teyssier} R.,  {Bouchet} F.~R.,  {Vibert} D.,   {Prunet} S.,
  2001, \mn@doi [\aap] {10.1051/0004-6361:20010692}, \href
  {https://ui.adsabs.harvard.edu/abs/2001A&A...374..358D} {374, 358}

\bibitem[\protect\citeauthoryear{Dwek \& Arendt}{Dwek \&
  Arendt}{1998}]{dwek1998tentative}
Dwek E.,  Arendt R.,  1998, The Astrophysical Journal, 508, L9

\bibitem[\protect\citeauthoryear{Dwek et~al.,}{Dwek
  et~al.}{1995}]{dwek1995morphology}
Dwek E.,  et~al., 1995, Astrophysical Journal, Part 1 (ISSN 0004-637X), vol.
  445, no. 2, p. 716-730, 445, 716

\bibitem[\protect\citeauthoryear{{Fine}, {Van Eck}  \& {Pratley}}{{Fine}
  et~al.}{2023}]{2023MNRAS.520.4822F}
{Fine} M.~A.,  {Van Eck} C.~L.,   {Pratley} L.,  2023, \mn@doi [\mnras]
  {10.1093/mnras/stad423}, \href
  {https://ui.adsabs.harvard.edu/abs/2023MNRAS.520.4822F} {520, 4822}

\bibitem[\protect\citeauthoryear{Finkbeiner, Davis  \& Schlegel}{Finkbeiner
  et~al.}{2000}]{finkbeiner2000detection}
Finkbeiner D.~P.,  Davis M.,   Schlegel D.~J.,  2000, The Astrophysical
  Journal, 544, 81

\bibitem[\protect\citeauthoryear{Fixsen, Cheng, Gales, Mather, Shafer  \&
  Wright}{Fixsen et~al.}{1996}]{fixsen1996cosmic}
Fixsen D.~J.,  Cheng E.,  Gales J.,  Mather J.~C.,  Shafer R.,   Wright E.,
  1996, The Astrophysical Journal, 473, 576

\bibitem[\protect\citeauthoryear{{Fixsen}, {Bennett}  \& {Mather}}{{Fixsen}
  et~al.}{1999}]{1999ApJ...526..207F}
{Fixsen} D.~J.,  {Bennett} C.~L.,   {Mather} J.~C.,  1999, \mn@doi [\apj]
  {10.1086/307962}, \href
  {https://ui.adsabs.harvard.edu/abs/1999ApJ...526..207F} {526, 207}

\bibitem[\protect\citeauthoryear{{Fixsen} et~al.,}{{Fixsen}
  et~al.}{2011}]{2011ApJ...734....5F}
{Fixsen} D.~J.,  et~al., 2011, \mn@doi [\apj] {10.1088/0004-637X/734/1/5},
  \href {https://ui.adsabs.harvard.edu/abs/2011ApJ...734....5F} {734, 5}

\bibitem[\protect\citeauthoryear{{G{\'o}rski}, {Hivon}, {Banday}, {Wandelt},
  {Hansen}, {Reinecke}  \& {Bartelmann}}{{G{\'o}rski} et~al.}{2005}]{healpix}
{G{\'o}rski} K.~M.,  {Hivon} E.,  {Banday} A.~J.,  {Wandelt} B.~D.,  {Hansen}
  F.~K.,  {Reinecke} M.,   {Bartelmann} M.,  2005, \mn@doi [\apj]
  {10.1086/427976}, \href
  {https://ui.adsabs.harvard.edu/abs/2005ApJ...622..759G} {622, 759}

\bibitem[\protect\citeauthoryear{Haslam, Salter, Stoffel  \& Wilson}{Haslam
  et~al.}{1982}]{haslam1982408}
Haslam C.,  Salter C.,  Stoffel H.,   Wilson W.,  1982, Astronomy and
  Astrophysics Supplement Series, vol. 47, Jan. 1982, p. 1, 2, 4-51, 53-142.,
  47, 1

\bibitem[\protect\citeauthoryear{{Hutschenreuter} et~al.,}{{Hutschenreuter}
  et~al.}{2022}]{2022A&A...657A..43H}
{Hutschenreuter} S.,  et~al., 2022, \mn@doi [\aap]
  {10.1051/0004-6361/202140486}, \href
  {https://ui.adsabs.harvard.edu/abs/2022A&A...657A..43H} {657, A43}

\bibitem[\protect\citeauthoryear{Ichiki}{Ichiki}{2014}]{ichiki2014cmb}
Ichiki K.,  2014, Progress of Theoretical and Experimental Physics, 2014,
  06B109

\bibitem[\protect\citeauthoryear{Janssen}{Janssen}{2020}]{Juno_Cruise_PDS}
Janssen M.,  2020, JUNO MWR CRUISE/SKY RDR DATA RECORDS V3.0,
  JNO-X-MWR-3-RDR-V3.0, \url{https://doi.org/10.17189/3hzx-jv21}

\bibitem[\protect\citeauthoryear{{Janssen} et~al.,}{{Janssen}
  et~al.}{2017}]{janssenetal}
{Janssen} M.~A.,  et~al., 2017, \mn@doi [\ssr] {10.1007/s11214-017-0349-5},
  \href {https://ui.adsabs.harvard.edu/abs/2017SSRv..213..139J} {213, 139}

\bibitem[\protect\citeauthoryear{{Keih{\"a}nen}, {Kurki-Suonio}, {Poutanen},
  {Maino}  \& {Burigana}}{{Keih{\"a}nen} et~al.}{2004}]{2004A&A...428..287K}
{Keih{\"a}nen} E.,  {Kurki-Suonio} H.,  {Poutanen} T.,  {Maino} D.,
  {Burigana} C.,  2004, \mn@doi [\aap] {10.1051/0004-6361:200400060}, \href
  {https://ui.adsabs.harvard.edu/abs/2004A&A...428..287K} {428, 287}

\bibitem[\protect\citeauthoryear{{Keih{\"a}nen}, {Kurki-Suonio}  \&
  {Poutanen}}{{Keih{\"a}nen} et~al.}{2005}]{2005MNRAS.360..390K}
{Keih{\"a}nen} E.,  {Kurki-Suonio} H.,   {Poutanen} T.,  2005, \mn@doi [\mnras]
  {10.1111/j.1365-2966.2005.09055.x}, \href
  {https://ui.adsabs.harvard.edu/abs/2005MNRAS.360..390K} {360, 390}

\bibitem[\protect\citeauthoryear{{Keskitalo} et~al.,}{{Keskitalo}
  et~al.}{2010}]{2010A&A...522A..94K}
{Keskitalo} R.,  et~al., 2010, \mn@doi [\aap] {10.1051/0004-6361/200912606},
  \href {https://ui.adsabs.harvard.edu/abs/2010A&A...522A..94K} {522, A94}

\bibitem[\protect\citeauthoryear{Kogut}{Kogut}{2012}]{kogut2012synchrotron}
Kogut A.,  2012, The Astrophysical Journal, 753, 110

\bibitem[\protect\citeauthoryear{Kogut et~al.,}{Kogut
  et~al.}{2011}]{kogut2011arcade}
Kogut A.,  et~al., 2011, The Astrophysical Journal, 734, 4

\bibitem[\protect\citeauthoryear{Lenz, Dor{\'e}  \& Lagache}{Lenz
  et~al.}{2019}]{lenz2019large}
Lenz D.,  Dor{\'e} O.,   Lagache G.,  2019, The Astrophysical Journal, 883, 75

\bibitem[\protect\citeauthoryear{Maeda, Alvarez, Aparici, May  \& Reich}{Maeda
  et~al.}{1999}]{maeda199945}
Maeda K.,  Alvarez H.,  Aparici J.,  May J.,   Reich P.,  1999, Astronomy and
  Astrophysics Supplement Series, 140, 145

\bibitem[\protect\citeauthoryear{{Mukai}, {Hansen}, {Mittskus}, {Taylor}  \&
  {Danos}}{{Mukai} et~al.}{2012}]{JunoPointingError}
{Mukai} R.,  {Hansen} D.,  {Mittskus} A.,  {Taylor} J.,   {Danos} M.,  2012,
  Juno Telecommunications,
  \url{https://descanso.jpl.nasa.gov/DPSummary/Juno_DESCANSO_Post121106H--Compact.pdf}

\bibitem[\protect\citeauthoryear{Orlando \& Strong}{Orlando \&
  Strong}{2013}]{orlando2013galactic}
Orlando E.,  Strong A.,  2013, Monthly Notices of the Royal Astronomical
  Society, 436, 2127

\bibitem[\protect\citeauthoryear{{Planck Collaboration} et~al.,}{{Planck
  Collaboration} et~al.}{2016}]{ade2016planck}
{Planck Collaboration} et~al., 2016, \mn@doi [\aap]
  {10.1051/0004-6361/201525830}, \href
  {https://ui.adsabs.harvard.edu/abs/2016A&A...594A..13P} {594, A13}

\bibitem[\protect\citeauthoryear{Poutanen et~al.,}{Poutanen
  et~al.}{2006}]{poutanen2006comparison}
Poutanen T.,  et~al., 2006, Astronomy \& Astrophysics, 449, 1311

\bibitem[\protect\citeauthoryear{Roger, Costain, Landecker  \& Swerdlyk}{Roger
  et~al.}{1999}]{roger1999radio}
Roger R.,  Costain C.,  Landecker T.,   Swerdlyk C.,  1999, Astronomy and
  Astrophysics Supplement Series, 137, 7

\bibitem[\protect\citeauthoryear{Rubi{\~n}o-Mart{\'\i}n
  et~al.,}{Rubi{\~n}o-Mart{\'\i}n et~al.}{2023}]{rubino2023quijote}
Rubi{\~n}o-Mart{\'\i}n J.,  et~al., 2023, Monthly Notices of the Royal
  Astronomical Society, 519, 3383

\bibitem[\protect\citeauthoryear{Serra, Lagache, Dor{\'e}, Pullen  \&
  White}{Serra et~al.}{2014}]{serra2014cross}
Serra P.,  Lagache G.,  Dor{\'e} O.,  Pullen A.,   White M.,  2014, Astronomy
  \& Astrophysics, 570, A98

\bibitem[\protect\citeauthoryear{Shaw, Sigurdson, Pen, Stebbins  \&
  Sitwell}{Shaw et~al.}{2014}]{shaw2014all}
Shaw J.~R.,  Sigurdson K.,  Pen U.-L.,  Stebbins A.,   Sitwell M.,  2014, The
  Astrophysical Journal, 781, 57

\bibitem[\protect\citeauthoryear{{Singal}, {Stawarz}, {Lawrence}  \&
  {Petrosian}}{{Singal} et~al.}{2010}]{2010MNRAS.409.1172S}
{Singal} J.,  {Stawarz} {\L}.,  {Lawrence} A.,   {Petrosian} V.,  2010, \mn@doi
  [\mnras] {10.1111/j.1365-2966.2010.17382.x}, \href
  {https://ui.adsabs.harvard.edu/abs/2010MNRAS.409.1172S} {409, 1172}

\bibitem[\protect\citeauthoryear{Smoot et~al.,}{Smoot
  et~al.}{1992}]{smoot1992structure}
Smoot G.~F.,  et~al., 1992, Astrophysical Journal, Part 2-Letters (ISSN
  0004-637X), vol. 396, no. 1, Sept. 1, 1992, p. L1-L5. Research supported by
  NASA., 396, L1

\bibitem[\protect\citeauthoryear{Spergel et~al.,}{Spergel
  et~al.}{2003}]{spergel2003first}
Spergel D.~N.,  et~al., 2003, The Astrophysical Journal Supplement Series, 148,
  175

\bibitem[\protect\citeauthoryear{{Thorne}, {Dunkley}, {Alonso}  \&
  {N{\ae}ss}}{{Thorne} et~al.}{2017}]{pysm}
{Thorne} B.,  {Dunkley} J.,  {Alonso} D.,   {N{\ae}ss} S.,  2017, \mn@doi
  [\mnras] {10.1093/mnras/stx949}, \href
  {https://ui.adsabs.harvard.edu/abs/2017MNRAS.469.2821T} {469, 2821}

\bibitem[\protect\citeauthoryear{Zheng et~al.,}{Zheng
  et~al.}{2017}]{zheng2017improved}
Zheng H.,  et~al., 2017, Monthly Notices of the Royal Astronomical Society,
  464, 3486

\bibitem[\protect\citeauthoryear{{Zonca}, {Singer}, {Lenz}, {Reinecke},
  {Rosset}, {Hivon}  \& {Gorski}}{{Zonca} et~al.}{2019}]{healpy}
{Zonca} A.,  {Singer} L.,  {Lenz} D.,  {Reinecke} M.,  {Rosset} C.,  {Hivon}
  E.,   {Gorski} K.,  2019, \mn@doi [The Journal of Open Source Software]
  {10.21105/joss.01298}, \href
  {https://ui.adsabs.harvard.edu/abs/2019JOSS....4.1298Z} {4, 1298}

\bibitem[\protect\citeauthoryear{de Oliveira-Costa, Tegmark, Gaensler, Jonas,
  Landecker  \& Reich}{de~Oliveira-Costa et~al.}{2008}]{de2008model}
de Oliveira-Costa A.,  Tegmark M.,  Gaensler B.,  Jonas J.,  Landecker T.,
  Reich P.,  2008, Monthly Notices of the Royal Astronomical Society, 388, 247

\makeatother
\end{thebibliography}
\bibliographystyle{mnras}

\end{document}